\documentclass[apj]{emulateapj}
\usepackage{apjfonts}
\usepackage{graphicx} 
\usepackage{amsmath} 
\usepackage{amssymb}
\usepackage{exscale, relsize}
\usepackage{amscd}
\usepackage{natbib}

\newcommand{\eg}{e.g.} 
\newcommand{\ie}{\textit{ie.}} 
\newcommand{\viz}{\textit{viz.}} 
\newcommand{\cf}{\textit{cf.}}

\newcommand{\sersic}{S\'ersic}

\newcommand{\model}{\texttt{model}}

\newcommand{\eff}{_\mathrm{e}}
\newcommand{\dyn}{_{\mathrm{d}, n}}

\newcommand{\obs}{_\mathrm{ob}}
\newcommand{\msimple}{\tilde{M}_\mathrm{d}}

\newcommand{\sol}{$_{\odot}$}
\newcommand{\dellogm}{M_*/M\dyn}

\begin{document}

\submitted{ApJ, accepted, 2010}
\title{On the Masses of Galaxies in the Local Universe}

\begin{abstract} 
   We compare estimates of stellar mass, $M_*$, and dynamical mass,
   $M_\mathrm{d}$, for a sample of galaxies from the Sloan Digital Sky
   Survey (SDSS). Under the assumption of dynamical homology (\ie ,
   $\msimple \sim \sigma_0^2 R\eff$, where $\sigma_0$ is the central
   velocity dispersion and $R\eff$ is the effective radius), we find a
   tight but strongly non-linear relation between the two mass
   estimates: the best fit relation is $M_* \propto \msimple^{0.73}$,
   with an observed scatter of 0.15 dex. We also find that, at fixed
   $M_*$, the ratio $M_*/\msimple$ depends strongly on galaxy
   structure, as parameterized by \sersic\ index, $n$. The size of the
   differential effect is on the order of 0.6 dex across $2 < n <
   10$. The apparent $n$-dependence of $M_*/\msimple$ is qualitatively
   and quantitatively similar to expectations from simple, spherical
   and isotropic dynamical models, indicating that assuming homology
   gives the wrong dynamical mass. To explore this possibility, we
   have also derived dynamical mass estimates that explicitly account
   for differences in galaxies' structures. Using this
   `structure-corrected' dynamical mass estimator, $M\dyn$, the best
   fit relation is $M_* \propto M\dyn^{0.92 \pm 0.01 (\pm 0.08)}$ with
   an observed scatter of 0.13 dex. While the data are thus consistent
   with a linear relation, they do prefer a slightly shallower
   slope. Further, we see only a small residual trend in $M_*/M\dyn$
   with $n$. We find no statistically significant systematic trends in
   $M_*/M\dyn$ as a function of observed quantities (\eg, apparent
   magnitude, redshift), or as a function of tracers of stellar
   populations (\eg, H$\alpha$ equivalent width, mean stellar age),
   nor do we find significantly different behavior for different kinds
   of galaxies (\ie, central versus satellite galaxies, emission
   versus non-emission galaxies). At 99 \% confidence, the net
   differential bias in $M_*/M\dyn$ across a wide range of stellar
   populations and star formation activities is $\lesssim 0.12$ dex
   ($\approx 40$ \%). The very good agreement between stellar mass and
   structure-corrected dynamical mass strongly suggests, but does not
   unambiguously prove, that:\ 1.)\ galaxy non-homology has a major
   impact on dynamical mass estimates, and 2.)\ there are not strong
   systematic biases in the stellar mass-to-light ratios derived from
   broadband optical SEDs.  Further, accepting the validity of both
   our stellar-- and dynamical--mass estimates, these results suggest
   that that the central dark--to--luminous mass ratio has a
   relatively weak mass dependence, but a very small scatter at fixed
   mass.
\end{abstract}

\author{Edward N Taylor$^{1, 2}$, Marijn Franx$^1$, Jarle
  Brinchmann$^1$, Arjen van der Wel$^3$, Pieter G. van Dokkum$^4$}

\affil{$^1$ Sterrewacht Leiden, Leiden University, 
  NL-2300 RA Leiden, Netherlands; ent@strw.leidenuniv.nl,\\
  $^2$ School of Physics, the University of Melbourne, 
  Parkville, 3010, Australia, \\
  $^3$ Max Planck Institut f\"ur Astronomie, D-69117 Heidelberg, Germany, \\
  $^4$ Department of Astronomy, Yale University, New Haven, CT
  06520-8101, USA}

\shorttitle{On the Masses of Galaxies in the Local Universe}
\shortauthors{Taylor, Franx, Brinchmann, van der Wel, van Dokkum}

%

\section{Introduction}

Techniques for estimating galaxies' stellar masses are a crucial tool for
understanding galaxies and their evolution. There are tight and
well-defined correlations between stellar mass and many other important
global properties like color, size, structure, metallicity, star
formation activity, and environment \citep[see, \eg,][]{Kauffmann2003b,
Shen2003, Blanton2005Env, Gallazzi2006}. Given a galaxy's stellar mass,
$M_*$, it is thus possible to predict a wide variety of global properties
with considerable accuracy. In this sense, stellar mass appears to be a
key parameter in determining (or at least describing) a galaxy's current
state of evolution. Moreover, since the growth of stellar mass (\cf\
absolute luminosity, color, etc.) is relatively slow and approximately
monotonic, stellar mass is a particularly useful parameter for
quantifying galaxy evolution.

Stellar mass estimates, whether derived from spectroscopic or photometric
spectral energy distributions (SEDs), are plagued by a variety of random
and systematic errors. These include a generic degeneracy between mean
stellar age, metallicity, and dust obscuration. It is typical to make the
simplifying assumptions that galaxies' stellar populations can be
described {\em en masse} (\ie, neglecting age/metallicity gradients and
complex dust geometries), and that galaxies' complex star formation
histories can be described parametrically. It is rare to attempt to
account for active galactic nucleus (AGN) emission. The stellar initial
mass function (IMF), including its universality or otherwise, remains a
major `known unknown'. Then there is the complication that different
wavelengths probe different aspects of the stellar population; the
inclusion of restframe UV or NIR data can thus, in principle and in
practice, have a large impact on the estimated stellar mass. These
effects are compounded by uncertainties in the stellar evolution models
themselves. A topical example is the importance of NIR-luminous thermally
pulsating asymptotic giant branch (TP-AGB) stars: for the same data and
stellar population parameters, the use of \citet{BruzualCharlot2003} or
\citet{Maraston2005} models can change the derived value of $M_*$ by a
factor of 3 for galaxies that host young ($\lesssim 1$ Gyr) stars, but
only if restframe NIR data are  included \citep{VanDerWel2006,
KannappanGawiser}. \citet{Conroy} have argued that the total random
uncertainties in $M_*$ are on the order of $\sim 0.3$ dex for galaxies at
$z \sim 0$. \looseness-1

For these reasons, it is essential to devise some way of assessing the
quality of stellar mass estimates through comparison to some other
fiducial mass estimate --- this is the primary motivation for the
present paper.  Specifically, using a number of the publicly available
`value added' catalogs of the Sloan Digital Sky Survey
\citep[SDSS;][]{York2000, Strauss2002}, we will compare stellar mass
estimates to total mass estimates derived from galaxy dynamics.

From the outset, we note that a difference between two quantities shows
only that: a difference. With no definitive standard to use as a basis
for comparison, the best that we can hope for is consistency between the
two mass estimates. Further, if and when there are differences, it is
impossible to unambiguously identify where the `fault' lies --- or even
if there is indeed a fault. For example, it is likely that the ratio
between stellar and total mass varies as a function of mass, and/or some
other global property/ies. We will also explore this issue in some
detail. \looseness-1

\vspace{0.2cm}

This kind of comparison has been done for SDSS galaxies by
\citet{DroryBenderHopp}, who considered both stellar mass estimates derived
from the SDSS spectra, as described by \citet{Kauffmann2003}, and those
derived using SED-fitting techniques that are commonly used at for
higher-redshift studies. These authors find a relatively tight correlation
between the two stellar mass estimates, with a mild systematic bias
depending on H$\alpha$ equivalent width (EW). This bias suggests a potential
problem with the stellar mass estimates as a function of specific star
formation rate (SSFR). Further, both stellar mass estimates correlated well with
the simple dynamical mass estimate, $\msimple$ (defined below), but showed
a clear trend in $M_*/\msimple$ with mass, such that less massive galaxies
had higher values of $M_*/\msimple$. \citet{Padman2004} used the mass
dependence of $M_*/\msimple$ to argue for an increasing stellar-to-dark
mass ratio for elliptical galaxies with higher masses, as did
\citet{Gallazzi2005}. Both \citet{Rettura2006} and \citet{VanDerWel2006}
have performed similar comparisons for $z \lesssim 1$ galaxies, with
similar conclusions.

\vspace{0.2cm}

It is common practice to derive a simple dynamical mass estimate based
on the velocity dispersion, $\sigma_0$, and the effective radius,
$R\eff$, via the scalar virial theorem:
\begin{equation}
  \mathrm{G} \msimple \approx k \sigma_0^2 R\eff ~ .
\label{eq:msimple} \end{equation}
(This is the dynamical mass estimator used for each of the studies
cited in the previous paragraph.)  The constant $k$ is usually assumed
to be in the range 3--5, and is intended to account for the `degree of
virialization', including the effects of dark matter and the intrinsic
shape of the velocity dispersion profile \citep[see,
\eg,][]{Cappellari2006, Gallazzi2006, VanDerWel2006}. By assuming a
constant $k$ for all galaxies, this expression implicitly assumes that all
galaxies are dynamically homologous, or
self-similar.\footnote{But see also \citet{WolfEtAl2009}, who derive a
mathematically identical relation from the spherical Jeans equation for a
system in dynamical equilibrium.}

But it is important to remember that the observed velocity dispersion is
actually the luminosity-weighted mean of the true, radially-dependent
velocity dispersion, projected onto the line of sight, and within the
spectroscopic aperture. The shape of mass profile has a strong influence
on the spatial and dynamical distribution of stellar orbits: in general,
the relation between the observed velocity dispersion and the underlying
mass profile thus depends on structure as well as size. As a dynamical
mass estimator, $\msimple$ can therefore only be considered as
approximate. (The tilde in $\msimple$ is intended to remind the reader of
this fact.)

\citet{Bertin2002} provide an analytic expression that makes it
possible to approximately account for this effect.  Using their
formulation of the problem, the dynamical mass can be expressed as:
\begin{equation} \label{eq:mdyn} 
\mathrm G M\dyn = K_V(n) \sigma_0^2 R\eff ~ ,
\end{equation}
Here, the term $K_V(n)$ encapsulates the effects of structure on stellar
dynamics.  (The subscript $n$ in $M\dyn$ is intended to make it clear
that non-homology has been accounted for as a function of \sersic\
index, $n$.)  For convenience, we will refer to $M\dyn$ as a
`structure corrected' dynamical mass estimator, but we
note that the inclusion of a structure-dependent term is not strictly
a correction.  \citet{Bertin2002} also provide an 
analytical approximation for $K_V(n)$: \looseness-1
\begin{equation} \label{eq:kvn}
K_V(n) \cong \frac{73.32}{10.465 + (n - 0.95)^2} + 0.954 ~ .
\end{equation}
This expression for $K_V(n)$ has been derived assuming a spherical
mass distribution that is dynamically isotropic and non-rotating, and
which, in projection, follows a \citet{Sersic1963, Sersic1968} surface
density profile.  For this (admittedly simple) scenario, this
approximate expression for $K_V(n)$ is accurate at the percent level
for $1 \le n \le 10$.  Substituting trial values of $n=2$ and $n=8$
into Equation \ref{eq:kvn} suggests that the differential effect of
non-homology on the inferred value of the dynamical mass is as much as
a factor of 3, or 0.5 dex.  Our first task in this paper, then, will be to
explore the importance of structural differences between galaxies,
using this prescription.

\vspace{0.2cm}

Before we begin, note that there are alternative approaches to exploring
the consistency between stellar and dynamical mass estimates. In
particular, a number of authors have considered the relation between
galaxies' stellar and dynamical masses in the context of well known
scaling relations between luminosity/mass and dynamics. For example,
\citet{BellDeJong} considered the relation between baryonic (\cf\
stellar) mass, $M_\mathrm{bar}$, and circular rotation velocity, $V_C$,
for disk galaxies --- the baryonic Tully-Fisher relation. These authors
showed that stellar mass estimates based on different passbands (\ie\
$M_*/L_V$ versus $M_*/L_K$) produced consistent $M_\mathrm{bar}$--$V_c$
relations. Furthermore, for a fixed IMF, they argued that it was possible
use a single color to estimate stellar mass-to-light ratios with an
accuracy of 0.1--0.2 dex.

There have also been a number of analogous studies for elliptical
galaxies, based on the fundamental plane \citep{FP1, FP2}, which can be
understood as a correlation between the dynamical mass-to-light ratio,
$M_\mathrm{d}/L$, and surface brightness. These studies (see, \eg,
Cappellari et al.\ 2006; and references therein) have tended to focus on
the `tilt' of the fundamental plane --- that is, the deviation of the
observed relation from the expectation assuming both a constant $M_*/L$
and structural and dynamical homology for all early type galaxies. The
tilt of the fundamental plane thus offers a means of probing variations
in $M_\mathrm{d}/L$ (including both variations in $M_*/L$ due to
different stellar populations, and variations in $M_*/M_\mathrm{d}$ due
to, \eg, different dark-to-luminous mass ratios) and/or the degree of
non/homology. While the relative importance of these different effects
remains an open question, it seems clear that, at least for early type
galaxies, both $M_*/L$ and $M_*/M_\mathrm{d}$ vary systematically with
mass \citep[see, \eg,][]{PrugnielSimien, Bertin2002, Trujillo2004,
Cappellari2006, LaBarbera2008, Allanson2009}.

\vspace{0.2cm}

This paper is structured as follows: in Section \ref{ch:data}, we
describe the various SDSS-derived catalogs that we will use, including
the definition of our galaxy sample. We validate the velocity dispersion
measurements used to derive dynamical masses in Appendix \ref{ch:faber}.
In Sections \ref{ch:results} and \ref{ch:results2}, we present two
parallel comparisons between stellar and dynamical mass estimates for
galaxies in our sample. First, in Section \ref{ch:results}, we use the
simple dynamical mass estimate $\msimple$; then, in Section
\ref{ch:results2} we show how the comparison changes using the
structure-corrected dynamical mass estimate, $M\dyn$. In Section
\ref{ch:consistency}, we explore the consistency between $M\dyn$ and
$M_*$. In particular, we will show that there are no statistically
significant trends in the ratio $M_*/M\dyn$ that would indicate
measurement biases in $M_*$ and/or $M\dyn$; this is not the case for the
simple dynamical mass estimate $\msimple$. We show in Appendix
\ref{ch:others} that these results are not unique to the sample we
consider in the main text. We discuss the interpretation and implications
of this result in Section \ref{ch:discussion}, before providing a summary
of our main results and conclusions in Section \ref{ch:summary}. \looseness-1

Throughout this work, we will assume the concordance cosmology; \viz.,
($\Omega_m,~ \Omega_\Lambda,~ \Omega_0$) = (0.3, 0.7, 1.0), and $H_0 =
70$ km/s/Mpc, and adopt a \citet{Chabrier2003} IMF.

\section{Data} \label{ch:data}

\defcitealias{Guo2009}{G09}

This work is based on data drawn from several publicly available
catalogs based on the SDSS dataset. Our analysis is based on redshifts
and velocity dispersions from the basic SDSS catalog for DR7
\citep{Abazajian2009}\footnote{Accessed via the Catalog Archive Server
  \citep[CAS; ][]{Thakar2008}: http://casjobs.sdss.org/CasJobs/}. We
use \sersic -fit structural parameters from \citet[hereafter
  \citetalias{Guo2009}]{Guo2009} and SED-fit stellar mass-to-light
ratio measurements from the DR7 Max-Planck-Institute for Astrophysics
(MPA)/Johns Hopkins University (JHU) value added
catalog\footnote{Available via
  http://www.mpa-garching.mpg.de/SDSS/}. In Appendix \ref{ch:others},
we repeat our analysis using the \sersic\ -fit structural parameters
given in the New York University (NYU) Value Added Galaxy Catalog
\citep[VAGC;][]{vagc} for DR7. Each of these catalogs have been well
described and documented in the references given; in this Section, we
only briefly summarize the most relevant aspects of each catalog for
the present work. \looseness-1

\subsection{Redshifts and Velocity Dispersions}

There are two sets of redshift and velocity dispersion measurements given
in the basic SDSS catalog for DR6 and DR7: the `spectro1D' values
produced by the Chicago group, and the `specBS' values produced by the
Princeton group. In terms of redshifts, the two algorithms produce
virtually identical results. The major difference between the two
algorithms is that, whereas the Chicago pipeline only gives velocity
dispersion measurements to those galaxies that are spectrally classified
as being `early type', all galaxies are given a velocity dispersion
measurement by the Princeton pipeline. From DR6, both the Princeton and
Chicago velocity dispersion measurement algorithms have been updated, so
as to eliminate the systematic bias at low dispersions identified by
\citet{Bernardi2007} for the DR5 values.\footnote{See
http://www.sdss.org/dr7/algorithms/veldisp.html for a discussion of the
spectro1D and specBS algorithms, as well as a comparison between these
values and those from \citet{Bernardi2003a,Bernardi2003b} and SDSS DR5.}
In Appendix \ref{ch:faber}, we compare both sets of velocity dispersions
to those given by \citet{Faber} for bright, early type galaxies: in both
cases, the values agree with the \citet{Faber} catalog values with an rms
difference of $\sim 18$ km/s and no discernible systematic
bias.\looseness-1

The default redshifts and velocity dispersions for in the SDSS catalog
(specifically, using SDSS parlance, the parameters \texttt{z} and
\texttt{veldisp} given in the table \texttt{specObjAll}) are the Chicago
values. For the sole reason that Princeton velocity dispersions are given
for all galaxies (rather than only the spectroscopically early-types) we
have elected to use the Princeton values instead; these are also the
measurements adopted for both the NYU and the MPA/JHU value added
catalogs.\looseness-1

\vspace{0.2cm}

As we have mentioned in the Introduction, the observed velocity
dispersion is the luminosity weighted average within the (projected)
spectroscopic aperture. In order to account for aperture effects, we have
scaled the observed value, $\sigma\obs$, for each galaxy to a central
velocity dispersion, $\sigma_0$, which is defined to be that that would
be observed within a circular aperture with a radius equal to 1/8 times
the apparent effective radius, $\Theta\eff$ \citet[see,
\eg][]{Jorgensen1995}. This correction has been made assuming $\sigma(R)
\propto R^{-0.066}$; \ie, $\sigma_0/\sigma\obs = (8
\Theta_\mathrm{ap}/\Theta\eff)^{-0.066}$, where $\Theta_\mathrm{ap} =
1\farcs5$ is the radius of the SDSS spectroscopic aperture. The scaling
of $\sigma(R)$ has been derived by taking a luminosity-weighted integral
of the spatially resolved velocity dispersions of galaxies from the
SAURON survey \citep{Cappellari2006}. The corrections themselves are
small --- the median correction is 0.02 dex, with an rms scatter of 0.02
dex --- and does not have a major impact on our results. Our qualitative
conclusions do not change if we assume the slightly weaker radial
dependence $\sigma(R) \propto R^{-0.04}$ as found by
\citet{Jorgensen1995}, or if we neglect this correction
altogether. \looseness-1

\vspace{0.2cm}

\subsection{\sersic\ Parameters: Size, Flux, and Structure}

\citetalias{Guo2009} have derived $r$-band structural parameters
including total magnitude, $m_\mathrm{tot}$, effective radius,
$\Theta\eff$, and \sersic\ index, $n$, for a modest sized sample of
SDSS galaxies.  (We discuss the specific sample selection in Section
\ref{ch:sample} below.)  These values have been derived via parametric
fits to the (2D) $r$-band surface brightness distribution of each
galaxy, assuming a \citet{Sersic1963,Sersic1968} profile, and
convolved with the appropriate PSF, using the publicly available code
\texttt{galfit} \citep{galfit}.  In order to account for blending,
where two galaxies are very close, both the target and companion(s)
are fit simultaneously.  Through analysis of simulated data,
\citetalias{Guo2009} show that the median error in each of
$m_\mathrm{tot}$, $\Theta\eff$, and $n$ to be less than 10
\%. (Although at the same time, they show that the uncertainties on
the derived parameters associated with background subtraction alone
can be significantly greater than this.)\looseness-1

In Appendix \ref{ch:others}, we will also make use of \sersic -fit
structural parameters from the NYU VAGC \citep{vagc}. Whereas
\texttt{galfit} considers the 2D surface brightness distribution, the VAGC
algorithm makes fits to the 1D azimuthally averaged curve of growth. The
analysis of simulated data presented by \citet{vagc} shows the VAGC
\sersic\ parameters to be systematically biased towards low fluxes, sizes,
and \sersic\ indices.  This problem becomes progressively worse for larger
$n$, such that sizes are underestimated by $\gtrsim 20$ \% and fluxes by
$\gtrsim 10$ \% for $n \gtrsim 5$. \citetalias{Guo2009} have shown that
this bias is produced by background over-estimation and over-subtraction
in the VAGC \sersic\ fits, owing to the use of a `local', rather than a
`global' background estimator. \looseness-1

\subsection{Stellar Masses} \label{ch:mstar}

We note that there are rather large differences between the the \sersic\
magnitudes given by \citetalias{Guo2009} and the default \model\
magnitudes given in the SDSS catalog. The \model\ photometry is derived
by making parametric fits to the 2D surface brightness distribution in
each band, using the sector fitting technique described by
\citet{Strauss2002}. These fits assume either an exponential or a De
Vaucouleurs profile; the profile shape is chosen based on the best-fit
$\chi^2$ in the $r$-band. For galaxies that are best fit by a De
Vaucouleurs \model\, we find $(m_{\mathrm{G09}, r}-m_{\mathrm{DeV}, r})
\approx -0.26 + 0.11 (n-4)$, where $n$ is the \sersic\ index reported by
G09; the scatter around this relation is at the level of 0.15 mag ($1
\sigma$). That is, even where G09 find $n = 4$, their flux is
approximately 0.26 mag brighter than the SDSS (De Vaucouleurs) \model\
flux; this discrepancy is larger for larger values of $n$. For this
reason, we take the $r$-band \sersic\ magnitude from \citetalias{Guo2009}
as a measure of total flux. \looseness-1

To derive a stellar mass, we then use $M_*/L$s taken from the MPA-JHU
catalog (DR7), which is maintained by the Garching
group.\footnote{Available via http://www.mpa-garching.mpg.de/SDSS/DR7/}
Note that, unlike previous MPA-JHU catalogs \citep[\eg][]{Kauffmann2003,
Brinchmann, Gallazzi2005}, which were based on the SDSS spectroscopy,
these masses are derived from fits to the $ugriz$ \model\
SEDs.\footnote{Note that in the SDSS algorithm, when deriving the $ugiz$
\model\ photometry, the structural parameters in the fits are held fixed
to the $r$-band values; only the overall normalization (\ie, total flux)
is allowed to vary. The fits in each band are also convolved with the
appropriate PSF. In this sense, the \model\ SEDs are both aperture
matched, and PSF-corrected.} Note, however, that the SED photometry has
been corrected for emission lines, according to the line--to--continuum
flux ratio in the spectroscopic fiber aperture (we discuss the importance
of this correction in Section \ref{ch:starpops} below). The SED fits are
based on the synthetic stellar population library described by
\citet{Gallazzi2005}, which have been constructed using the
\citet{BruzualCharlot2003} stellar population models, assuming a
\citet{Chabrier2003} stellar IMF. These $M_*/L$s have been shown to be in
good agreement (rms in $\Delta \log M_*$ of $\lesssim 0.1$ dex) with the
spectrally-derived values described by \citet{Kauffmann2003} for
DR4.\footnote{See
http://www.mpa-garching.mpg.de/SDSS/DR7/mass\_comp.html}

\subsection{Sample Definition} \label{ch:sample}

Our decision to rely on the \sersic\ structural parameters derived by
\citetalias{Guo2009} restricts us to working with their sample.  Our
rationale for this decision is that, in our estimation, the G09
\sersic\ fits are the most robust that are presently available. \looseness-1

The G09 sample was constructed with the specific goal of exploring
differences in the sizes and structures of `central' and `satellite'
galaxies in groups and clusters. To this end, they selected 911 $z < 0.08$
`centrals' as the first-ranked (in terms of $M_*$) group/cluster members
from the \citet{Yang2007} group catalog, which was in turn constructed
from the DR4 NYU VAGC. These galaxies were selected to have a flat
logarithmic distribution in halo mass in the range $11.85 < \log
M_\mathrm{halo}/$M\sol\ $ < 13.85$ (800 galaxies), plus 100 galaxies in
the range $13.85 < \log M_\mathrm{halo}/$ M\sol\ $ < 14.35$, and all 11
central galaxies in clusters with $\log M_\mathrm{halo}/$M\sol\ $>14.35$.
In this way, the central galaxy sample was constructed to span a
representative range of (large) halo masses.

G09 also construct two $z < 0.08$ `satellite' control samples, in which
the satellite galaxies are selected to match the central galaxies. For
the first of these, satellites are chosen to match centrals in $M_*$ to
within 0.08 dex; in the second, satellites are also required to match
centrals to within 0.03 mag in $^{0.1}(g-i)$ color. Because more massive
galaxies are more likely to be (counted as) centrals, not every central
has a satellite counterpart within these limits: the matching is more
than 90 \% successful for $M_* < 10^{10.85}$ M\sol, and less than 10 \%
successful for $M_* > 10^{11.15}$ M\sol. The two satellite samples, so
constructed, consist of 769 and 746 galaxies, respectively.

G09 exclude a number of these galaxies from their analysis because of
confusion, leaving a sample of 879 central galaxies, and two samples
of 704 and 696 satellites each.  While duplicates are not allowed
within the individual satellite samples, some galaxies do appear in
both samples; combining the two satellite samples we have 1167 unique
galaxies.  We exclude a further 71 galaxies whose spectra are not
deemed `science worthy' by the SDSS team (\ie\ the flag
\texttt{sciencePrimary} is set to zero).  In order to avoid very large
errors in the dynamical mass estimates, we also exclude 160 galaxies
that have relative errors in their velocity dispersion measurements
that are greater than 10 \%.  This requirement excludes mostly low-$n$
and low-$M_*$ galaxies: the vast majority of these 160 galaxies have
$M_* <10^{10.8}$ M\sol\ and $n < 1.5$.  Our results do not depend on
these selections.  We are thus left with a sample of 1816 galaxies, of
which 784 have been selected as central galaxies, and 1032 have been
selected as being satellites of comparable mass. \looseness-1

\vspace{0.2cm}

The major disadvantage to using the G09 sample is that the relative
number of central/satellite galaxies is not at all representative of the
general galaxy population. However it is worth noting that G09 have shown
that, at least for structurally early type galaxies, and after matching
both color and mass, there are no differences in the sizes and structures
of central and satellite galaxies. They thus conclude that the
distinction between central and satellite has no impact on galaxy
structure. This already suggests that the G09 sample may be adequately
statistically representative of the massive galaxy population.
\looseness-1

Even so, we will explicitly examine the possible role of sample selection
effects in shaping our results in Section \ref{ch:bias} by comparing
different subsamples from within the combined G09 sample. Further, in
Appendix \ref{ch:others}, we analyze a more general galaxy sample, using
structural parameters from either the NYU VAGC \sersic\ fits or the SDSS
De Vaucouleurs/exponential \model\ fits.

\section[Comparing $M_*$ and $\msimple$ Assuming Dynamical Homology]
{Results I --- Comparing Stellar and Dynamical Mass Estimates Assuming Dynamical Homology} 
\label{ch:results}

In this Section and the next, we present parallel comparisons between
stellar mass and two different estimates of dynamical mass. As we have
said in the Introduction, it is common practice to obtain a simple
dynamical mass estimate based on $\sigma_0$ and $R\eff$ alone, using the
scalar virial theorem; \viz\ $\msimple \approx k \sigma_0^2 R\eff$. In
Section \ref{ch:raw}, we directly compare the values of $M_*$ and
$\msimple$ for the \citetalias{Guo2009} galaxies; we will assume $k = 4$.
We will then argue in Section \ref{ch:structure} that the agreement
between stellar and dynamical mass estimates may be significantly
improved if we allow for non-homology. To test this idea, in Section
\ref{ch:results2}, we will perform the same comparisons using the
structure corrected dynamical mass estimator, $M\dyn$.

\subsection{The Relation Between Stellar and Dynamical Mass} 
\label{ch:raw}

\begin{figure} \centering 
\includegraphics[width=8.8cm]{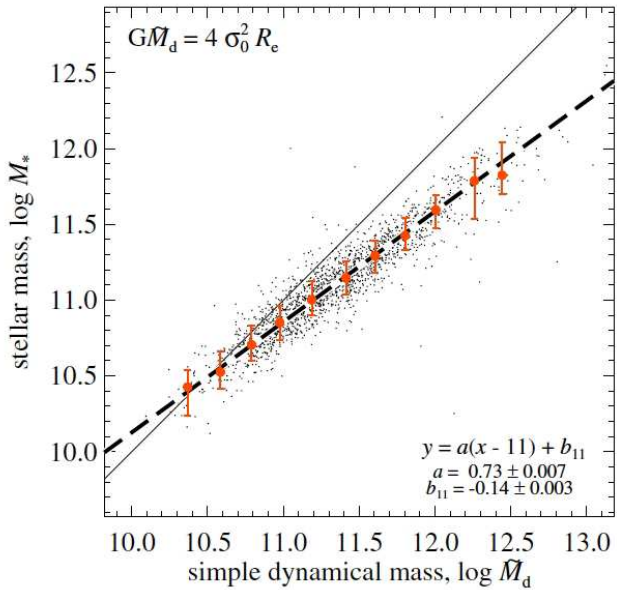}
\caption{Comparing stellar and dynamical mass estimates under the
  assumption of dynamical homology. --- This Figure shows the relation between
  stellar mass and a simple estimate of dynamical mass,
  $\mathrm{G}\msimple = 4 \sigma_0^2 R\eff$, for galaxies in the
  \citetalias{Guo2009} sample.  The black points show the data themselves;
  the red points with error bars show the median and 16/84 percentile
  values of $M_*$ in narrow bins of $\msimple$; the heavy dashed line
  shows a log-linear fit to the data, with the form and parameters as
  given.  {\em While there is a relatively tight correspondence
  between $M_*$ and $\msimple$, the relation is clearly non-linear}.
  Further, for $\msimple \lesssim 10^{10.5}$ M\sol, $M_*$ appears to exceed $\msimple$ for the majority of galaxies, which is logically
  inconsistent.  We explore these results in greater detail in Figures
  \ref{fig:rawsides} and \ref{fig:rawcond}.  In Figure
  \ref{fig:corrd}, we show how these results change if we account for
  structural and dynamical non-homology in our estimates of dynamical mass.
\label{fig:raw}}
\end{figure}
`
In \textbf{Figure \ref{fig:raw}}, we compare the values of the simple
dynamical mass estimator, $\msimple$, to the values of $M_*$ for galaxies
in the G09 sample. The first thing to notice is that there is a
relatively tight but clearly nonlinear relation between $\msimple$ and
$M_*$, such that $M_* \propto \msimple^a$ with $a < 1$. Moreover, this
simple analysis suggests that for many galaxies, including the majority
of galaxies with $\msimple \lesssim 10^{10.5}$ M\sol, $M_*$ actually {\em
exceeds} $\msimple$. This is logically inconsistent, and necessarily
implies a problem in the calculation of $M_*$ and/or $\msimple$.

\begin{figure*} \centering
\includegraphics[width=17.8cm]{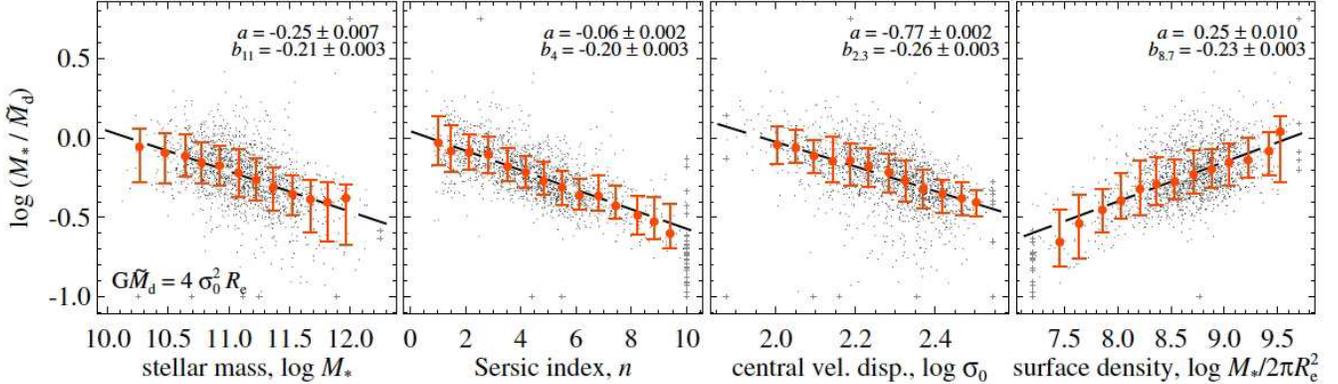}
\caption{Comparing stellar and dynamical mass estimates under the
  assumption of dynamical homology. --- Each panel of this Figure plots the
  stellar-to-dynamical mass ratio, $M_*/\msimple$, as a function of
  (left to right) mass, structure, velocity dispersion, and effective
  surface density.  Within each panel, the black points show the data
  themselves; points that fall outside the plotted range are shown
  with a small grey plus; the large points with error bars show the
  median and 16/84 percentiles of $\log(M_*/\msimple)$ in narrow bins
  of the quantity shown on the $x$-axis.  The dashed lines show fits
  to the data of the form $y = a(x - X) + b_X$, in analogy to Figure
  \ref{fig:raw}.  {\em At least when using this (overly) simple way of
  estimating galaxies' dynamical masses, there are strong trends in
  $M_*/\msimple$ with both mass and structure}.  We see similarly
  tight and strong trends with velocity dispersion and effective
  surface density, as well as with other parameters like size (not
  shown).
  \label{fig:rawsides}}
\end{figure*}
\begin{figure*} \centering
\includegraphics[width=17.8cm]{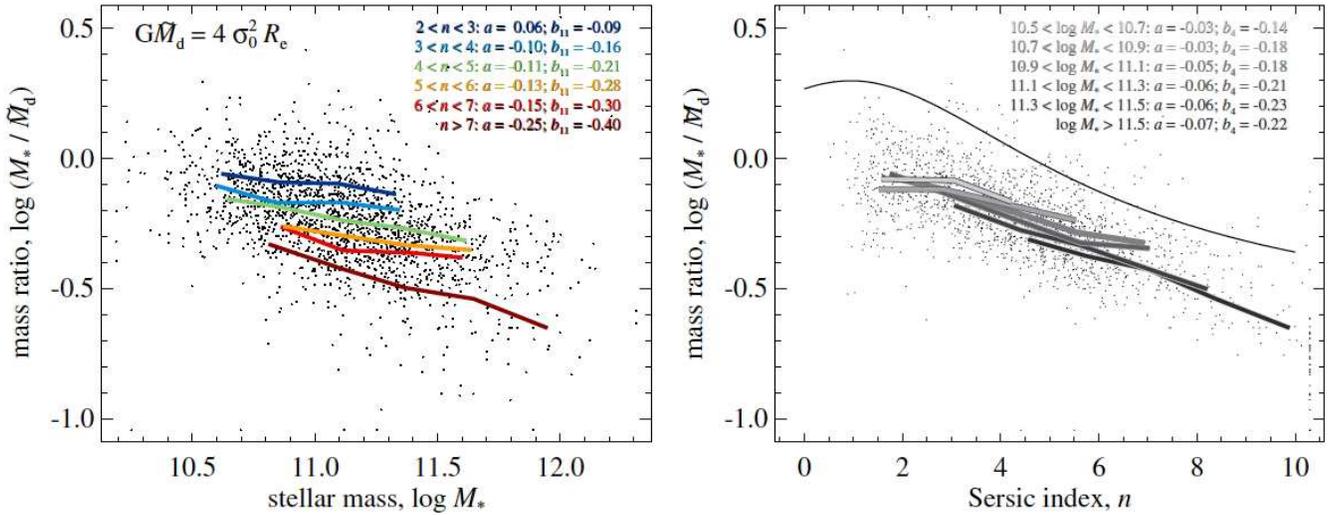}
\caption{Separating out the mass- and structure-dependences of the
  stellar-to-dynamical mass ratio, $M_*/\msimple$. --- In the left
  panel, we plot $M_*/\msimple$ as a function of $M_*$; the thick
  lines in this panel show the median relation in bins of $n$.  In the
  right panel, we do the reverse: $M_*/\msimple$ is plotted as a
  function of $n$, with the solid lines showing the median relation in
  bins of $M_*$.  In both panels, we give the parameters of the best
  fit relation for galaxies in each of the bins shown.  It is clear
  from both panels that {\em at fixed $M_*$, the scatter in
  $M_*/\msimple$ is directly related to $n$.}  It is also true that,
  even at fixed $n$, $M_*/\msimple$ appears to vary with $M_*$; we
  explore this issue further in Figure \ref{fig:corrcond}.  For this
  Figure, we have assumed $\mathrm{G}\msimple = 4 \sigma_0^2 R\eff$;
  in the right-hand panel, the solid curve shows the expected shape 
  of the $M_*/\msimple$ relation for a spherical and dynamically
  isotropic system that follows a \sersic\ profile, derived using
  Equation \ref{eq:kvn}.  The general agreement between the shape
  of this curve and that of the $M_*/\msimple$--$n$ relation suggests
  that including this term may significantly improve the
  correspondence between stellar and dynamical mass estimates. We
  explore this issue further in Figures \ref{fig:corrd} and
  \ref{fig:corrsides}.
  \label{fig:rawcond}}
\end{figure*} 

Before discussing this result further, it is appropriate to make a few
comments about the random errors in our estimates of $M_*$ and
$\msimple$. In particular, it is important to realize that the errors in
$\Theta\eff$, $m_\mathrm{tot}$, and $n$ are strongly covariant: for
example, an error in the structural index will affect the values of both
$\Theta\eff$ and $m_\mathrm{tot}$. Because $M_*$ depends on the measured
value of $m_\mathrm{tot}$, and $\msimple$ on the measured value of
$\Theta\eff$, $M_*$ and $\msimple$ are thus also covariant. This makes
the seemingly trivial task of fitting a line to the observed
$M_*$--$\msimple$ relation rather problematic. To do this properly would
require full and consistent treatment of the covariant uncertainties in
the \sersic -fit parameters, but this information is not given by
\citetalias{Guo2009}. \looseness-1

Our solution to this problem is simply to minimize the mean absolute
perpendicular distance between the data and the fit. When doing so, we
also use a `sigma-clipping' algorithm to avoid the influence of the most
egregious outliers; specifically, we iteratively exclude points that lie
off the best-fit line by more than 5 times the rms offset. While the
gradient of the best-fit line does depend on the fitting scheme used (we
will explore this in more detail in Section \ref{ch:corrd}), the best fit
parameters are not strongly dependent on how aggressively we sigma-clip.
In order to avoid strong covariances between the slope and intercept of
the best-fit line, we actually compute the fit in terms of $\log(
\msimple/10^{11} $M\sol); that is, we fit a relation of the form $y = a(x
-11) + b_{11}$. Statistical uncertainties on the fit parameters have been
derived from bootstrap resampling.
The best fit to the $M_*$--$\msimple$ relation, so derived, is shown
as line heavy dashed line in Figure \ref{fig:raw}.  The best fit
parameters are $a = 0.73 \pm 0.007$ and $b_{11} = -0.14 \pm 0.003$.

In \textbf{Figure \ref{fig:rawsides}}, we explore the relation between
$M_*$ and $\msimple$ in greater detail. The different panels of this
Figure show the difference between $M_*$ and $\msimple$ as a function of
several interesting global properties: (from left to right) galaxy mass,
structure, dynamics, and surface density. It clear that that
$M_*/\msimple$ is strongly correlated with all four of these parameters.
For each of the parameters shown, the size of the median trend in
$M_*/\msimple$ across the sample is on the order of 0.5 dex, although it
is slightly lower for $M_*$ and slightly higher for effective surface
density. \looseness-1

To quantify this statement, we have again made fits to the data,
assuming the form $y = a(x - X) + b_X$, where $X$ is an arbitrary
value chosen to be close to the median value of the quantity $x$ for
our sample.  For these fits, in contrast to the previous Section, we
have minimized the mean absolute {\em vertical} offset between the
data and the fit.  Again, we use a non-aggressive sigma-clipping
scheme to exclude extreme outliers.  (In all that follows, when
considering the stellar-to-dynamical mass ratio, we will always fit in
this way; we will only use the minimum perpendicular distance
algorithm described above when fitting the relation between stellar
and dynamical masses.)  The best fit lines to the data, so derived,
are shown as the heavy dashed lines in each panel; the best-fit
parameters are given in each panel.  In the case of \sersic\ index,
the scatter around the best fit relation is $\approx 0.12$ dex; for
the other three parameters it is slightly higher: $\approx 0.15$ dex.\looseness-1

\begin{figure} \centering
\includegraphics[width=8.8cm]{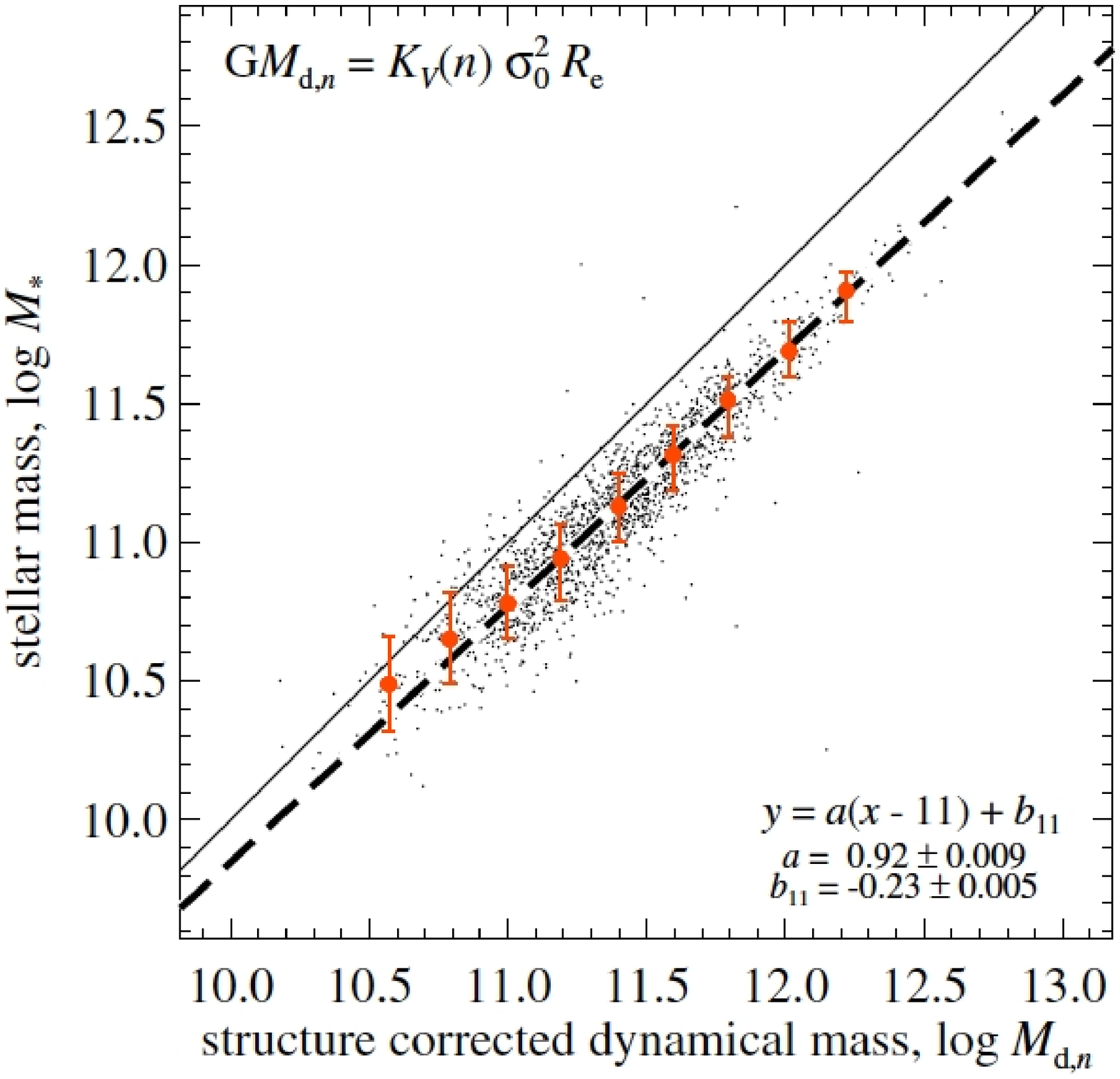}
\caption{Comparing stellar and dynamical mass estimates, accounting
  for both structural and dynamical non-homology. --- The difference between
  this Figure and Figure \ref{fig:raw} is that we have used $\mathrm{G}
  M\dyn = K_V(n) \sigma_0^2 R\eff^2$, with $K_V(n)$ defined as in
  Equation \ref{eq:kvn}; otherwise all symbols and their meanings are
  as in Figure \ref{fig:raw}.  The key point to be made from this
  Figure, in comparison to Figure \ref{fig:raw}, is that {\em allowing
  for non-homology makes a big difference to the inferred dynamical
  masses}, and so to the correspondence between stellar and dynamical
  masses. Further, we note that the apparent inconsistency whereby $M_*
  > \msimple$ for $M_* \lesssim 10^{10.5}$ M\sol\ galaxies seen in
  Figure \ref{fig:raw} is no longer apparent.  After accounting for structure-dependent differences in
  galaxies' dynamics, the relation between $M_*$ and $M\dyn$ is much
  more nearly linear.  However it remains true that the difference
  between stellar and dynamical mass appears to grow with increasing
  mass.  \label{fig:corrd} \looseness-1}
\end{figure}

\subsection{The Importance of Galaxy Structure \\ in Dynamical Mass Estimates} \label{ch:structure}

There are of course strong correlations between mass, velocity
dispersion, surface density, and structure. It is thus possible that the
apparent trend with any given parameter in Figure \ref{fig:rawsides} is
`spurious', in the sense that it is driven by a trend in another more
`fundamental' parameter. We note that galaxies' star formation activity
and histories have been shown to correlate closely with all of mass,
velocity dispersion, and surface density \citep[see,
\eg,][]{Kauffmann2003, Kauffmann2003b, Kauffmann2006, Franx2008,
Graves2009}. Indeed, with the assumption that $M_* \sim M\dyn$, these
three quantities are all related by factors of $R\eff$, which is also
closely correlated with $M_*$ \citep{Shen2003, Franx2008}. But the fact
that $M_*/\msimple$ depends on galaxy \emph{structure} --- and
particularly the agreement between the observed trend and expectations
derived from a simple dynamical model --- immediately suggests that
structure-dependent differences in galaxy dynamics may play a role in the
results shown in Figure \ref{fig:raw}. With this in mind, in
\textbf{Figure \ref{fig:rawcond}} we attempt to separate out the $M_*$--
and $n$--dependences of $M_*/\msimple$. Specifically, we want to test the
hypothesis that departure from linearity in the $M_*$--$\msimple$
relation seen in Figure \ref{fig:raw} is at least in part a function of
structure, and not mass. \looseness-1


Figure \ref{fig:rawcond}a shows $M_*/\msimple$ as a function of $M_*$;
the colored lines show the median relation in bins of \sersic\ index. The
median relation between $M_*/\msimple$ and $M_*$ has a rather similar
slope for each different $n$ bin: $M_*/\msimple$ does depend on mass. If
the dynamical mass-to-light ratio were to depend on mass only, however,
we would expect the relations for different \sersic\ indices to overlap.
Instead, the relations for each bin are clearly offset from one another.
That is, at fixed mass, the scatter in $M_*/\msimple$ is closely
correlated with galaxy structure. \looseness-1

\begin{figure*}
\centering \includegraphics[width=17.8cm]{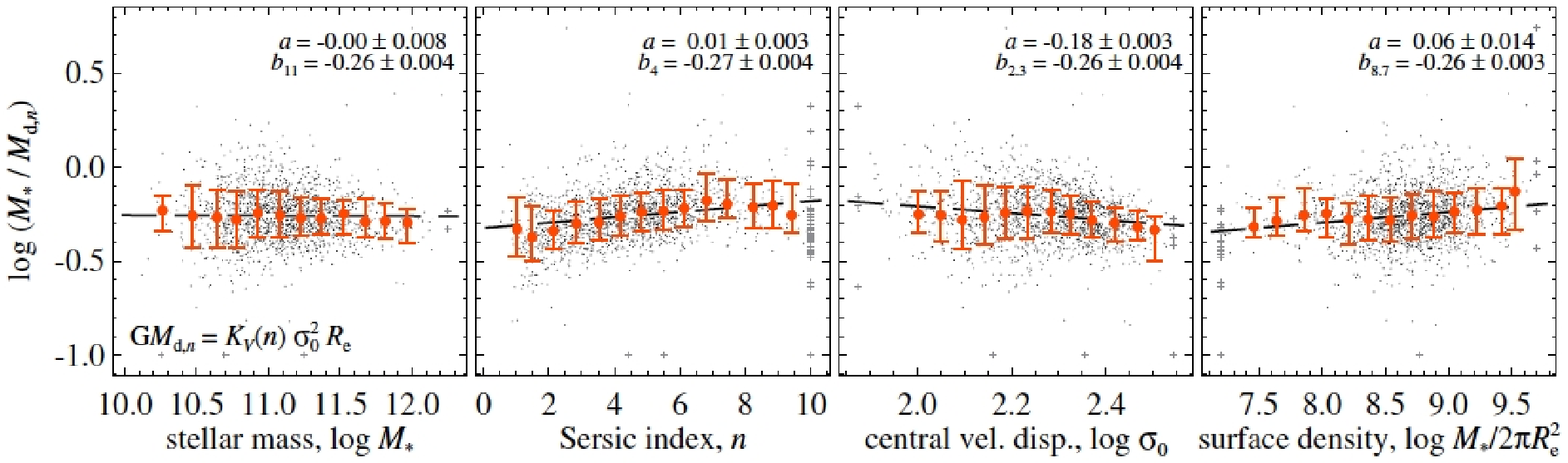}
\caption{Comparing stellar and dynamical mass estimates accounting for
  structure-dependent differences in galaxy dynamics. --- The
  difference between this Figure and Figure \ref{fig:rawsides} is that
  we have used $\mathrm{G} M\dyn = K_V(n) \sigma_0^2 R\eff^2$, with
  $K_V(n)$ defined as in Equation \ref{eq:kvn}; otherwise all symbols
  and their meanings as is in Figure \ref{fig:raw}.  {\em After
  accounting for structure dependent differences in galaxy dynamics,
  the apparent trends in $\dellogm$ with stellar mass and \sersic\
  index are substantially reduced}.  The apparent trends with other
  properties, including velocity dispersion, surface density, size,
  and color, are also substantially reduced, or effectively disappear
  (see also Figures \ref{fig:bias} and \ref{fig:starpops}).
  \label{fig:corrsides}}
\end{figure*}
\begin{figure*} \centering
\vspace{0.6cm}
\includegraphics[width=17.8cm]{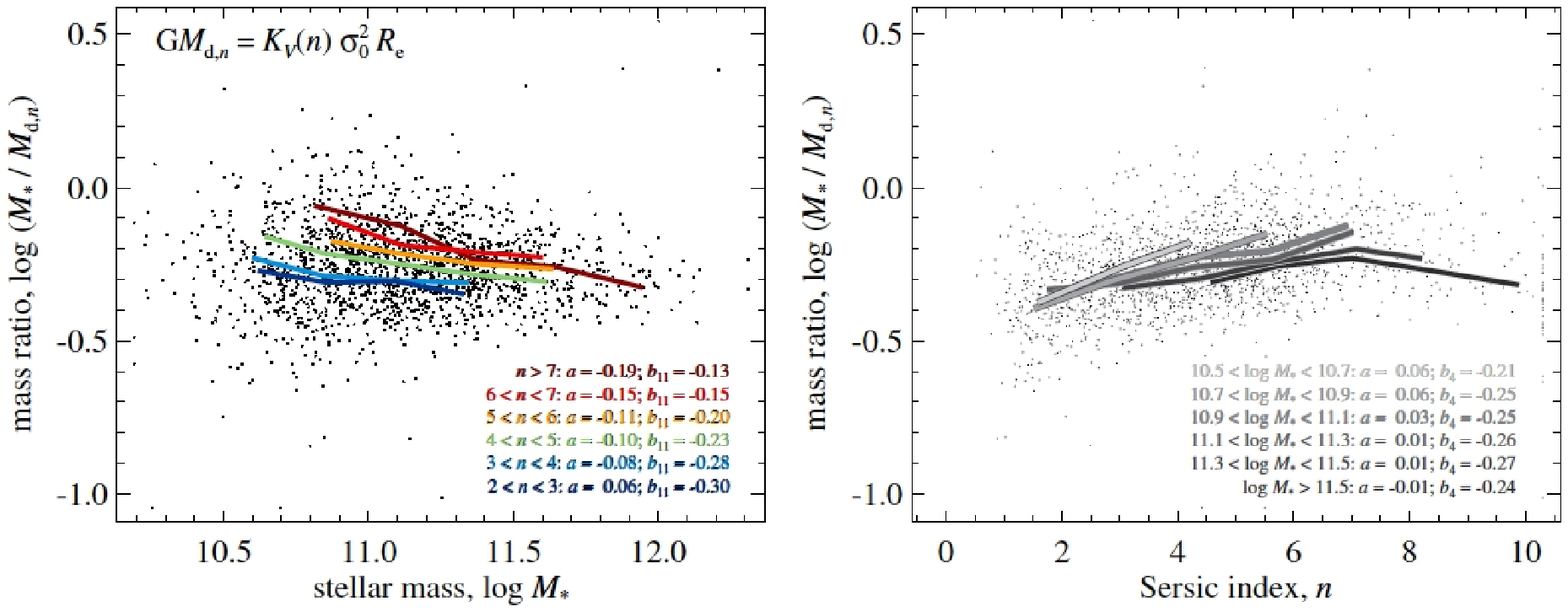}
\caption{Does $\dellogm$ vary with $M_*$, $n$, or both? --- The
  difference between this Figure and Figure \ref{fig:rawcond} is that
  we have accounted for structure-dependent differences in galaxy
  dynamics through the $K_V(n)$ term defined in Equation \ref{eq:kvn};
  otherwise all symbols and their meanings as is in Figure
  \ref{fig:rawcond}.  As in Figure \ref{fig:rawcond}, it is clear that
  the scatter in $\dellogm$ at fixed $M_*$ depends strongly on $n$;
  conversely, at fixed $n$, there is also a strong correlation
  between $M_*$ and $\dellogm$.  This effect appears to be stronger
  for higher values of $n$ and $M_*$.  In comparison to Figure
  \ref{fig:rawcond}, the inclusion of an $n$-dependent dynamical term
  obviously changes the slope of the $\dellogm$--$n$ relation, but
  cannot affect the slope of the $\dellogm$--$M_*$ relation at fixed
  $n$.  In other words, the apparent $n$-dependence of $\dellogm$ at
  fixed $M_*$ is sensitive to the specific model used to derive
  $K_V(n)$; on the other hand, the result that, at fixed $n$,
  $\dellogm$ varies with $M_*$ is insensitive assumed form of
  $K_V(n)$.  \label{fig:corrcond}}
\end{figure*} 

In Figure \ref{fig:rawcond}b, we do the opposite: in this panel, we plot
$M_*/\msimple$ as a function of \sersic\ index, and the different lines
show the median relation in bins of stellar mass. Again, it is clear that
$M_*/\msimple$ depends on both $M_*$ and $n$: the median relations for
each different mass bin are roughly parallel, but offset from one another.\looseness-1

Further, the different mass bins in Figure \ref{fig:rawcond}b cover
different ranges in \sersic\ index: where the lowest mass bin is
dominated by galaxies with $1 < n < 4$, the highest mass bin is
dominated by $n > 4$ galaxies.  Similar behavior can be seen in Figure
\ref{fig:rawcond}a: the lowest $n$ bin contains very few galaxies
with $\log M_* > 11.3$ M\sol , and virtually all $\log M_* > 11.5$
M\sol\ galaxies are in the $n > 7$ bin.  That is, there is a
correlation between $M_*$ and $n$ within our sample.  Because the
trend is towards lower values of $M_*/\msimple$ with increasing $n$,
this correlation contributes to the apparent mass dependence of the
dynamical-to-stellar mass ratio.

\vspace{0.2cm}

Again, the dynamical mass estimates used thus far have been derived under
the assumption of dynamical homology (\ie\ $K_V$ is equal to 4 for all
galaxies). The solid line in Figure \ref{fig:rawcond}b shows the expected shape of the relation between $M_*/\msimple$ and $n$, derived from Equation \ref{eq:kvn}. The agreement between the observed relation between
$M_*/\msimple$ and $n$ and the expectations from this (admittedly simple)
dynamical model immediately suggests that non-homology has an important
effect on dynamical mass estimates. We note that both
\citet{PrugnielSimien} and \citet{Trujillo2004} have made a similar
argument for the importance of non-homology based on the fundamental plane
of elliptical galaxies (see also Section \ref{ch:discussion} below).
\looseness-1

\section[Comparing $M_*$ and $M\dyn$ Accounting for Non-Homology]
{Results II.--- Comparing Stellar and Dynamical Mass Estimates Accounting for Dynamical Non-Homology} 
\label{ch:results2}

In this Section, we investigate the potential impact of non-homology on
the agreement between stellar and dynamical mass estimates. To this end,
we repeat the comparisons between stellar and dynamical mass estimates
presented above, using the structure corrected dynamical mass estimator,
$M\dyn$, in place of the simple estimate $\msimple$. We quantify the
relation between $M_*$ and $M\dyn$ for our sample in Section
\ref{ch:corrd}, and show in Section \ref{ch:mstarn} that allowing for
non-homology considerably improves the correspondence between stellar and
dynamical mass estimates. \looseness-1

\subsection{The Relation Between Stellar and Dynamical Mass} \label{ch:corrd}
 
In \textbf{Figure \ref{fig:corrd}}, we show the relation between stellar
and dynamical mass for the \citetalias{Guo2009} sample, using the
structure corrected dynamical mass estimator, $M\dyn$; this Figure should
be compared to Figure \ref{fig:raw}. It is immediately obvious that the
correlation between $M_*$ and $M\dyn$ is much closer to linear than that
between $M_*$ and the simple dynamical mass, $\msimple$. Further, we note
that the results are now logically consistent, in that $M_* < M\dyn$ for
almost all galaxies. This is our most basic result: {\em
structure-dependent differences in galaxy dynamics can have a big impact
on the inferred dynamical mass}, and so the stellar-to-dynamical mass
ratio. 

\vspace{0.2cm}

The best-fit parameters for the $M_*$--$M\dyn$ relation are $a = 0.92 \pm
0.007$ and $b_{11} = -0.23 \pm 0.004$, where we have used bootstrap
resampling to estimate the statistical uncertainty. While the statistical
errors in the fit parameters are impressively small, systematic errors are
sure to dominate. To see this, consider what would happen if we were to
impose a stellar mass limit $M_* > M_\mathrm{lim}$ in Figure
\ref{fig:corrd}: we would only include those galaxies with $M\dyn <
M_\mathrm{lim}$ that have high values of $M_*/M\dyn$; similarly, we would
exclude those galaxies with $M\dyn < M_\mathrm{lim}$ that have low values
of $M_*/M\dyn$. This would lead to a significantly shallower best-fit
slope to the $M_*/M\dyn$ relation. As a specific example, if we were only
to consider galaxies from the \citetalias{Guo2009} sample with $M_* >
10^{10.8}$ M\sol, we would find $M_* \propto M\dyn^{0.86}$. In this context, it is worth noting both that the scatter in the $M_*$--$M\dyn$ relation appears appears to increase with decreasing mass, and that the lowest mass points in Figure \ref{fig:corrd} lie systematically above the best fit line.

Note that,
while we have phrased this in terms of a mass selection effect, at least
part of this effect is related to how we have fit for the slope of the
$M_*$--$M\dyn$ relation; \viz, by minimizing the mean perpendicular
distance between the data and the best-fit line. If we were instead to fit
by minimizing the mean offset in $M_*/M\dyn$ as a function of $M_*$, we
would reduce our vulnerability to this effect. \looseness-1

There are thus two effects that have the potential to systematically bias
the measured slope of the $M_*$--$M\dyn$ relation. We can obtain a simple
estimate for the systematic error on the parameters $a$ and $b_{11}$ by
re-fitting the $M_*$--$M\dyn$ relations in different ways. If we fit by
minimizing the mean vertical offset, $\Delta M_*$, from the best-fit
$M_*/M\dyn$ relation, we find $a = 0.85$ and $b_{11} = -0.20$. If instead
we fit by minimizing the mean horizontal offset, $\Delta M\dyn$, we find
$a = 1.00$ and $b_{11} = -0.26$. That is, the systematic uncertainties
related to the technique used to fit the $M_*$--$M\dyn$ relation (for this
sample) are on the order $\Delta a = 0.08$ and $\Delta b_{11} = 0.03$.
\looseness-1

\vspace{0.2cm}

What about the systematic biases due to the particular mass distribution
of galaxies in the \citetalias{Guo2009} sample? To explore the importance
of these effects in our measurement of the slope of the $M_*$--$M\dyn$
relation, we have tried re-fitting the $M_*$--$M\dyn$ relation, weighting
each point according to its stellar mass. The specific weights have been
derived through a comparison between the mass distribution of galaxies
within the \citetalias{Guo2009} sample, in bins of $\Delta M_* = 0.1$ dex,
and the $z \sim 0$ mass function of \citet{Bell2003}. We have chosen the
weight for each galaxies so that that the weighted stellar mass
distribution of the \citetalias{Guo2009} sample matches the `real' stellar
mass function. This weighting scheme is akin to $1/V_\mathrm{max}$
weighting, inasmuch as if one were able to derive $V_\mathrm{max}$ values
for the sample, one would hope to obtain similar values.

Re-fitting the \citetalias{Guo2009} sample using these weights, we find $a
= {1.00 \pm 0.05}$. The larger random error on this value in comparison to
our fiducial values stems from the fact that the lower mass galaxies are
given much greater weights (by several orders of magnitude); the
inclusion/exclusion of these points in the bootstrap resampling thus has a
major impact on the best-fit slope. The fact that using these weightings
gives an almost perfectly linear relation between $M_*$ and $M\dyn$ is
striking, but it is important to remember that this fit is based almost
entirely on the relatively small number of $M_* \lesssim 10^{11}$ M\sol\
galaxies in the sample. In Appendix \ref{ch:others}, we perform the same
analysis for a more general galaxy sample, drawn from the NYU VAGC, which
provides a much better sampling of the true galaxy mass function. Using
the weighting scheme described above, the relative weights of galaxies
with $10.2 < \log M_* / $M\sol$ < 11.7$ differ by a factor of only 10. For
this galaxy sample, we find $a = 0.930 \pm 0.004\ (^{+0.03}_{-0.07})$
without weighting, compared to $a = 0.933 \pm 0.007$ with weighting. \looseness-1

While the data are consistent with a linear relation between $M_*$ and
$M\dyn$, they thus prefer a slightly shallower relation. For the
\citetalias{Guo2009} sample, we find $a = 0.93 \pm 0.007 \ (\pm 0.07)$,
and $b_{11} = -0.23 \pm 0.004\, (\pm 0.03)$. This should be compared to
the values of $a = 0.73 \pm 0.006\ (^{+0.07}_{-0.03})$ and $b_{11} = -0.14
\pm 0.003\ (^{+0.01}_{-0.03})$ for the simple dynamical mass estimate,
$\msimple$. While the mass distribution of galaxies within the
\citetalias{Guo2009} sample can in principle induce a large bias in the
measured slope of the $M_*$--$M\dyn$ relation, our best fit value is in
fact consistent with that derived from a more general galaxy sample, in
which these effects play a far smaller role. We will explore the potential
role of other sample selection effects in Section \ref{ch:selfx}.

\vspace{0.2cm}

In \textbf{Figure \ref{fig:corrsides}}, we show the trends in $M_*/M\dyn$
with mass, structure, dynamics, and density; this Figure should be
compared to Figure \ref{fig:rawsides}. For each of these four parameters,
the trends in $M_*/M\dyn$ are significantly weaker than what we have seen
for $M_*/\msimple$. The net differential trend across the sample is now on
the order of 0.2 dex or so, as compared to 0.5 dex for $M_*/\msimple$.
While there is still a strongly statistically significant trend in
$M_*/M\dyn$ with $\sigma$, the trend with surface density is now only
significant at the $4 \sigma$ level. While we do still see signs of a
trend in $M_*/M\dyn$ with $n$, this trend is not statistically
significant, at least for the sample as a whole. We discuss this point
further in the next Section.

\subsection{Does $M_*/M\dyn$ Depend on Mass, or Structure, or Both?}
\label{ch:mstarn}

In \textbf{Figure \ref{fig:corrcond}}, we return to the issue of the $n$-
and $M_*$-dependence of $M_*/M\dyn$; this Figure should be compared to
Figure \ref{fig:rawcond}. In Figure \ref{fig:corrcond}a, we show the
median relation between $M_*/M\dyn$ and $M_*$ in bins of $n$. Again, the
trends in $M_*/M\dyn$ with $M_*$ for the different $n$-bins are parallel,
but offset from one another. In each of the $3 \lesssim n \lesssim 7$
bins, we find that $M_*/M\dyn$ scales approximately as $M_*^{0.1}$; that
is, roughly consistent with the scaling that we see for the sample as a
whole. Figure \ref{fig:corrcond}b shows the median relation between
$M_*/M\dyn$ as a function of $n$ for different bins in $M_*$. While the
trend in $M_*/M\dyn$ with $n$ is substantially weaker than we saw using
the simple dynamical mass, $\msimple$, we still see that $M_*/M\dyn$
varies with $n$; if anything, it would appear that by using the
prescription for $K_V(n)$ given in Equation \ref{eq:kvn}, we have
overcorrected for the effects of non-homology. Without detailed dynamical
modeling, however, we have no means of refining the model used to derive
Equation \ref{eq:kvn}. (We will discuss this point further in Section
\ref{ch:discussion}.)

\vspace{0.2cm}

In other words, {\em we have shown that accounting for structural and
dynamical homology significantly improves the agreement between
stellar and dynamical mass estimates as a function of \sersic\ index,
n,} but we have not unambiguously shown whether or not $M_*/M\dyn$
depends on galaxy structure --- nor can we.

\section[Exploring Potential Biases in $M_*/M\dyn$]
{Results III --- Exploring Potential Biases in $M_*/M\dyn$}
\label{ch:consistency}

In this Section we discuss three general classes of biases that may
affect the results we have presented in Section \ref{ch:results2}:
first, systematic biases in the \sersic\ fits that we use to derive
$M_*$ and $M\dyn$ (Section \ref{ch:bias}; see also Appendix
\ref{ch:others}); then, the possibility of severe selection effects
for the \citetalias{Guo2009} sample (Section \ref{ch:selfx}; see also
Appendix \ref{ch:others}); and finally, systematic effects associated
with the estimation of stellar mass-to-light ratios (Section
\ref{ch:starpops}).  We will show very good consistency between the
values of $M_*$ and $M\dyn$; this is not the case for the simple
estimate $\msimple$.  As in the previous Section, this consistency
provides strong circumstantial evidence --- but not proof beyond a
reasonable doubt --- that there are no significant biases in either
measurement. \looseness-1

\subsection{Looking for Possible Observational Biases} \label{ch:bias}

\begin{figure*} \centering
\includegraphics[width=17.8cm]{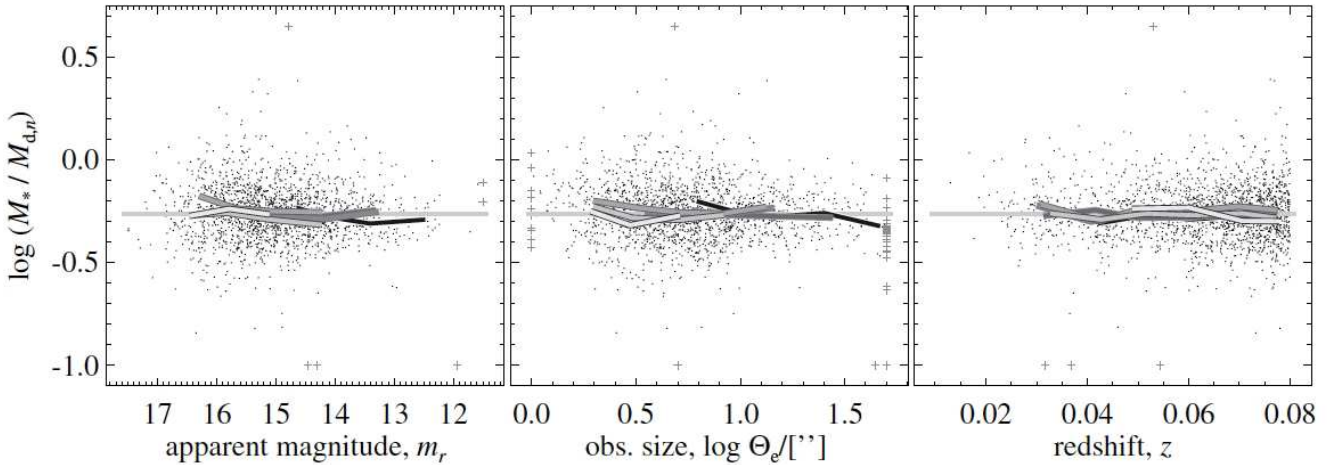}
\caption{Signs of observational biases? --- Each panel shows $\dellogm$
  a function of a direct observable.  Within each panel, the solid
  lines show the median relation in bins of stellar mass; these bins
  are the same as those shown in the right-hand panels of Figures
  \ref{fig:rawcond} and \ref{fig:corrcond}.  The points show the data
  themselves; points that fall outside the range of each panel are
  shown as a small grey plus.  Within the \citetalias{Guo2009} sample,
  there are correlations between \sersic\ index and each of the
  observed quantities shown in this Figure; however, there are only
  very weak trends in $\dellogm$ with any of these observables.
  Moreover, each of the stellar mass bins follows essentially the same
  median relation. This suggests that {\em neither the stellar nor
  dynamical mass estimates are obviously seriously biased by
  systematic errors in the \sersic\ fits}.
  \label{fig:bias}}
\end{figure*}

We explore the possibility of serious observational biases in
\textbf{Figure \ref{fig:bias}}. In each panel of this Figure, we plot
$M_*/M\dyn$ as a function of a basic observable: namely (left to right),
apparent magnitude, apparent size, and redshift. The lines in each panel
show the median relation for the same bins of stellar mass as are shown
in Figures \ref{fig:rawcond}b and \ref{fig:corrcond}b. \looseness-1

By a similar argument to the one given in Section \ref{ch:mstarn}, if
the apparent mass dependence of the ratio $M_*/M\dyn$ were driven by
observational effects that are strong functions of apparent brightness
or size, then we would expect there to be clear trends in $M_*/M\dyn$
for each of the stellar mass bins in Figure \ref{fig:bias}.  This is
not obviously the case.  Fitting to the data in each stellar mass bin,
the gradients of the best-fit relation between $M_*/M\dyn$ and
all of apparent magnitude, apparent size, and redshift are consistent
with zero; this is true for each stellar mass bin individually
(typically within $1 \sigma$), as well as for the sample as a whole
(within 1.2--$1.4 \sigma$). \looseness-1

We note that the same is not true using, for example, the \sersic\
structural parameters given in the NYU VAGC to derive $M_*$ and
$M\dyn$ (as we do in Appendix \ref{ch:others}).  In this case, we do
see a weakly statistically significant gradient (at the level of $4
\sigma$) with observed size.  That is, this kind of test is indeed
able to (weakly) detect mild systematic errors in the \sersic -fit
parameters on the order of 10--20 \%.  The results in Figure
\ref{fig:bias} thus argue against the idea that there are any serious
biases affecting the measurement of $M_*$ or $M\dyn$ (or, more
accurately, the ratio $M_*/M\dyn$) associated with the \sersic -fit
structural parameters used to derive these values.\looseness-1

\vspace{0.2cm}

As can be seen in Figure \ref{fig:corrsides}, there is a statistically
significant correlation between $M_*/M\dyn$ and velocity dispersion,
$\sigma$.  Using the same argument as above, it is conceivable that
this could be produced by a systematic bias in the measured values of
$\sigma$.  The observed trend of $\sim -0.12$ dex in $M_*/M\dyn$ over
$\sim 0.5$ dex in $\sigma$ could be entirely explained by a $\sim
0.06$ dex differential bias in the measured values of $\sigma$.  In
connection with this point, we stress that a comparison between the
SDSS measured values of $\sigma$ and those of \citet{Faber} shows no
systematic biases.  This should give some confidence that the trend in
$M_*/M\dyn$ with $\sigma$ is real, and not a product of observational
biases.

\subsection{Looking for Potential Sample Selection Effects} \label{ch:selfx}

\begin{figure*} \centering
\includegraphics[width=17cm]{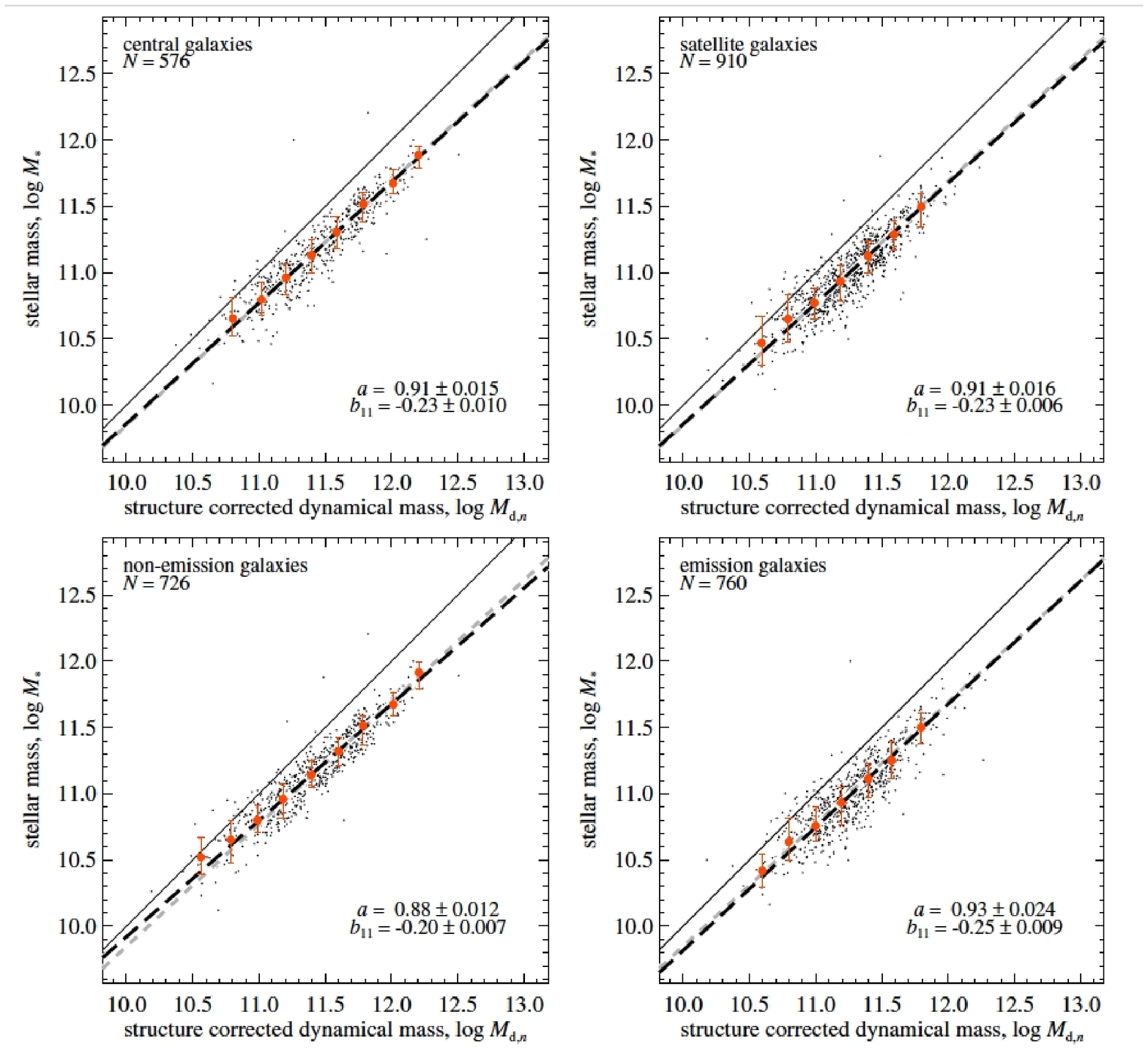}
\caption{Possible sample selection effects? --- The
  \citetalias{Guo2009} sample is not necessarily representative of the
  general galaxy population, in terms of the distribution of masses,
  environments, or star formation activity.  To explore the potential
  role of these effects, each panel of this Figure shows the
  $M_*$--$M\dyn$ comparison for different subsets of the
  \citetalias{Guo2009} sample, distinguishing between central and
  satellite galaxies (upper panels) and between spectroscopically
  emission and non-emission galaxies (lower panels).  For the lower
  panels, the spectroscopic classification is based on the BPT
  diagram, following the scheme of \citet{Brinchmann}; the `emission'
  sample includes both AGN-dominated and composite spectra galaxies.
  In the upper panels, we do not show those satellite/central that do
  not have \citet{Brinchmann} spectral classifications; that is, the
  same samples are plotted in both the upper and the lower panels.
  The dashed grey line in each panel shows the best fit relation for
  the entire \citetalias{Guo2009}. {\em We find a consistent
    $M_*$--$M\dyn$ relation for all four of these subsamples, as well
    as the \citetalias{Guo2009} as a whole}.  While there is possibly
  a slight offset in $\dellogm$ between star-forming and passive
  galaxies, we note that this disappears if we only consider $n > 2$
  galaxies.  That is, this offset appears to be due to the different
  $n$-distributions of the star-forming and passive samples, rather
  than an intrinsic difference between the values of $M_*/M\dyn$ for
  emission and non-emission galaxies (see also Figure
  \ref{fig:starpops}).  We also note that for each of the subsamples
  shown, within statistical errors, we find consistent behavior in
  $\dellogm$ at fixed $M_*$ and $n$ as is shown in Figure
  \ref{fig:corrcond}. We therefore conclude that selection effects do
  not play a major role in shaping our results (see also Appendix
  \ref{ch:others}).
\label{fig:subsamples}}
\end{figure*}

As we have described in Section \ref{ch:sample}, the \citetalias{Guo2009}
sample has been constructed in such a way that massive galaxies in
general, and in particular central galaxies in very massive halos, are
drastically overrepresented in comparison to the general field
population.  If there are systematic differences in $M_*/M\dyn$ as a
function of, for example, environment or star formation activity,
there is thus a very real danger that sample selection effects may
play an important role in shaping our results.

We explore this issue in \textbf{Figure \ref{fig:subsamples}}, in
which we have divided the \citetalias{Guo2009} sample in
central/satellite and non-/emission subsamples, in order to look for
differences between these populations. Here, we have selected
`non-emission' galaxies as H$\alpha$ and H$\beta$ in absorption.  The
`emission' subsample contains both star forming galaxies and AGN
hosts. \looseness-1

In the upper panels of Figure \ref{fig:subsamples}, it is clear that
we find essentially identical relations between $M_*$ and $M\dyn$ for
the central and satellite galaxy subsamples.  Given that, as we have
shown in Figure \ref{fig:corrcond}, $M_*$/$M\dyn$ varies with both
$M_*$ and $n$, it is not all that surprising that central and
satellite galaxies show the same $M_*$--$M\dyn$ relation: not only
have the two subsamples have been constructed to be matched in $M_*$,
\citetalias{Guo2009} have shown that there are no structural differences
between satellites and centrals at fixed mass. \looseness-1

However, we do find very slightly different $M_*$--$M\dyn$ relations
for the emission and non-emission subsamples. Formally, the two
relations are slightly offset from one another, at the level of 0.05
dex; the gradients of the two relations differ at the $2 \sigma$
level. That said, for the $10.5 < \log M_* < 12$ range spanned by the
sample, the difference between the best-fit relations for each
subsample and that for the sample as a whole are $\lesssim 0.03$
dex. We also note that the apparent offset disappears (at least within
$2 \sigma$) if we consider only the $n > 2$ galaxies in both
subsamples. Moreover, we do not find any differences in $M_*/M\dyn$ at
fixed $M_*$ and $n$ between these different subsamples --- within
statistical uncertainties, each subsample shows the same behavior as
is seen in Figure \ref{fig:corrcond}. This suggests that the apparent
offset between for the emission and non-emission galaxies is driven by
the different distributions of \sersic\ indices within the emission
and non-emission subsamples, rather than any difference in
$M_*$/$M\dyn$ intrinsically related to galaxy activity. (See also
Section \ref{ch:starpops}). \looseness-1

\vspace{0.2cm}

That is, our results do not appear to be strongly influenced by the
relative numbers of central/satellite galaxies or of non-/emission
galaxies in our sample.  The above analysis does suggest, however,
that the measured relation between $M_*$ and $M\dyn$ is sensitive to
the joint $M_*$--$n$ distribution within the sample.  This is a direct
consequence of the fact that $M_*/M\dyn$ depends on both mass and
\sersic\ index (Figure \ref{fig:corrcond}).  We have considered biases
associated with the mass distribution in Section \ref{ch:corrd}.  But
note that if, at fixed mass, the distribution of $n$ within the
\citetalias{Guo2009} sample differs significantly from the `true'
distribution for a general field population, then we may therefore
find a very different slope for the $M_*/M\dyn$ relation.
\looseness-1

For this reason, in Appendix \ref{ch:others}, we repeat our $M_*$--$M\dyn$
comparisons for a more general sample of $0.035 < z < 0.08$ field
galaxies, using the structural parameters given in the NYU
VAGC.\footnote{As we have already remarked, the VAGC \sersic\ fits have
been shown to suffer from systematic errors arising from background
oversubtraction. But, as we also show in Appendix \ref{ch:others}, it
turns out that the ratio $M_*$/$M\dyn$ is extremely robust to random or
systematic errors in the fitting of structural parameters, so long as the
apparent magnitude, effective radius, and \sersic\ index are derived
consistently, and the term $K_V(n)$ is included in the definition of the
dynamical mass (See also Section \ref{ch:discussion}). Further, we find
very similar $M_*$--$M\dyn$ relations for the \citetalias{Guo2009} sample
using either VAGC or the \citetalias{Guo2009} values for the structural
parameters. Any large differences in the measured $M_*$-$M\dyn$ relation
between the \citetalias{Guo2009} sample and the general field sample would
therefore necessarily be a product of selection effects.} The best fit
logarithmic slope of the $M_*$--$M\dyn$ relation for $0.035 < z < 0.08$
field galaxies is $a = 0.91 \pm 0.003$, compared to $a = 0.92 \pm 0.009$
for the \citetalias{Guo2009} sample. This suggests that the
\citetalias{Guo2009} sample is not grossly biased in terms of the
distribution of $n$ at fixed mass. (Here again, it is significant that
\citetalias{Guo2009} have found that, at fixed mass, there are no
structural differences between satellite and central galaxies.)

To summarize the results of this section, then, separate analysis of
central/ satellite and non-/emission galaxies suggest that our results
are not strongly affected by selection effects associated with these
properties.  Furthermore, although the observed slope of the
$M_*/M\dyn$ relation is in principle sensitive to the joint $M_*$--$n$
distribution within the sample, we find very little difference between
the \citetalias{Guo2009} sample and a more general field galaxy sample.
Finally, we stress that we do not find any evidence that selection
effects have an important impact on the results shown in Figure
\ref{fig:corrcond}; \ie, the observation that the ratio
$M_*/M\dyn$ depends on both $n$ (at fixed $M_*$) and on $M_*$ (at
fixed $n$).

\subsection{Looking for Biases in the Stellar Mass-to-Light Ratio Estimates}
\label{ch:starpops}

\begin{figure*} 
\includegraphics[width=17.cm]{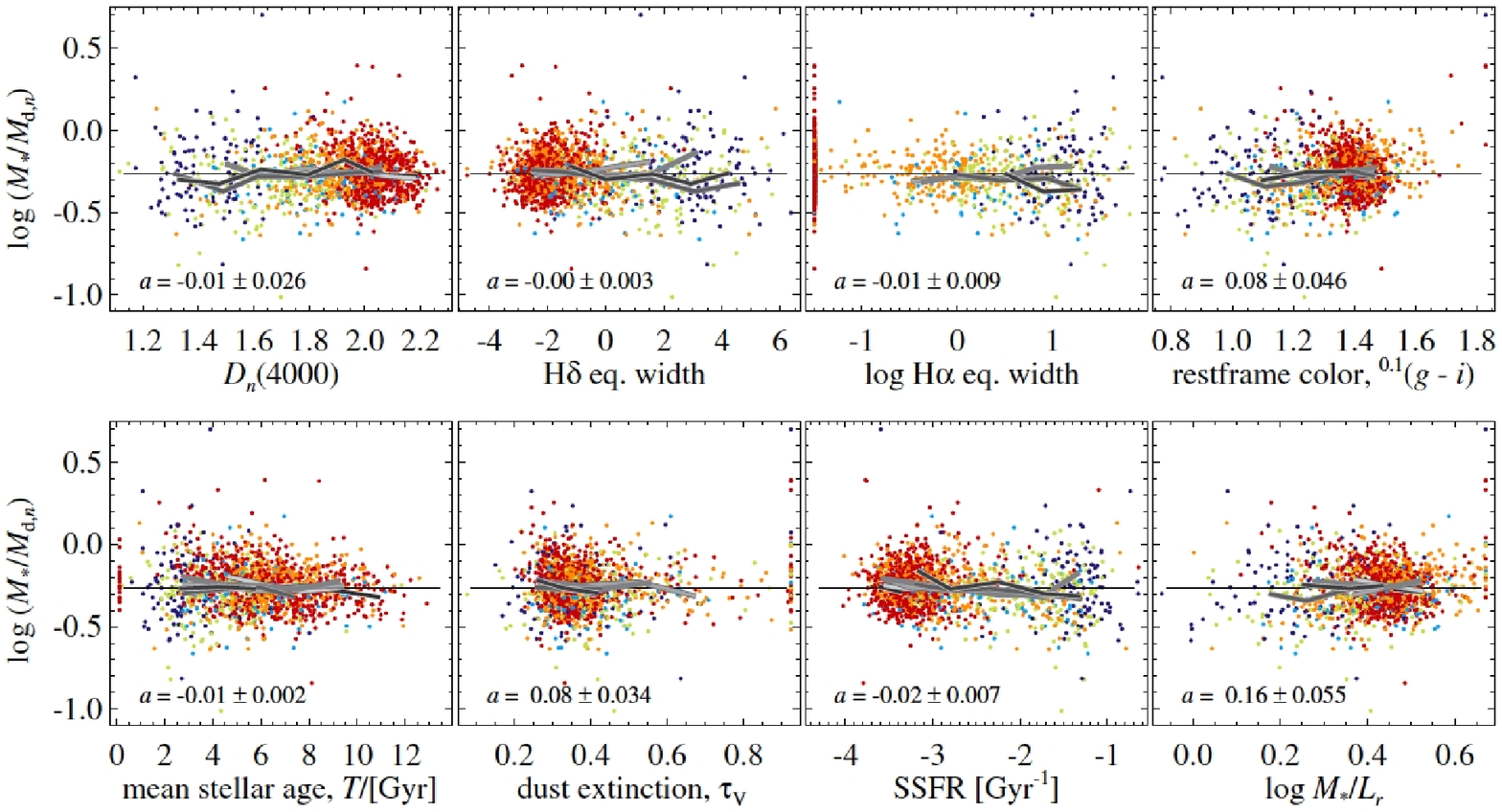}
\caption{Stellar population-dependent effects? --- Each panel plots
  $\dellogm$ as a function of a stellar population diagnostic.  The
  upper panels show directly observed quantities; the lower panels
  show stellar population parameters derived from synthetic stellar
  population modeling.  In these panels, the mean stellar age shown
  is that given by \citet{Kauffmann2003}, which are based on spectra,
  and are thus formally independent of $M_*/L$.  The dust extinction,
  specific star formation rates, and $M_*/L$s are all
  self-consistently derived from the same SED fits, as described in
  Section \ref{ch:mstar}.  Within each panel, individual points are
  color-coded according to spectral classifications; specifically:
  star forming (blue), low S/N star forming (cyan), composite
  (yellow), AGN-dominated (orange), and `unclassifiable' (\ie\,
  non-emission; red) galaxies.  Within each panel, the black points
  with error bars show the median and 16/84 percentiles of $\dellogm$
  in bins.  We see no statistically significant systematic
  differences in $M_*/M\dyn$ for galaxies with different stellar populations or
  star formation histories.  This argues against there being major
  problems with the stellar population models used to estimate
  $M_*/L$.  At 99 \% confidence, these results suggest that any differential biases as a function of the
  parameters shown is $\lesssim 0.12$ dex ($\approx 40$ \%).
\label{fig:starpops}}
\end{figure*}

As we have stated in the Introduction, our primary motivation for
comparing stellar and dynamical mass estimates is to validate the
stellar mass estimates.  We explore this issue in \textbf{Figure
\ref{fig:starpops}}.  In each panel of this Figure, we plot the ratio
$M_*/M\dyn$ as a function of a different property of the stellar
population.  The solid lines in each panel show the median relation
for the same stellar mass bins shown in Figures \ref{fig:rawcond} and
\ref{fig:corrcond}.  The upper panels plot $M_*/M\dyn$ as a function
of a direct observable; the lower panels plot $M_*/M\dyn$ as a
function of a derived property.  Note that the stellar mass estimates
we have used were derived from the $ugriz$ photometry, rather than
spectra.  The measured values of $M_*/M\dyn$ are thus formally
independent of the three spectral measurements shown in the top
panels.  Further, note that the age estimate that we show is taken
from \citet{Kauffmann2003}; these values are also derived from the
spectra.  The dust obscuration and specific star formation rates
(SSFRs) shown come from the SED fits used to derive $M_*/L$; these
values are thus self-consistently derived.

We have color-coded the data in Figure \ref{fig:starpops} according to
their spectral classification as given by \citet{Brinchmann}; \viz:
star forming (blue); low S:N star forming (cyan); composite (yellow),
AGN (orange), and non-emission (red). In general, we see little if any
differences in the values of $M_*/M\dyn$ for different stellar
populations.  There are not obviously large differences between the mean
values of $M_*/M\dyn$ for AGN hosts, star forming galaxies, or
non-emission galaxies.  

Within each panel we give the gradient of the best-fit line for the
whole sample.  These values are all statistically consistent with
zero.  The only possible exceptions to this rule are the gradients in
$M_*/M\dyn$ as a function of H$\alpha$ equivalent width (EW) and as a
function of $M_*/L_r$, both of which are non-zero at the $\sim
3\sigma$ level.  Just as there is little if any trend for the sample
as a whole, there are no statistically significant trends for any of
the individual mass bins.

We can quantify the degree of correspondence between stellar and
dynamical mass estimates by considering the differential bias between
galaxies over the range of each of the properties shown in Figure
\ref{fig:starpops}.  Looking at the median relations shown for each
mass bin suggests that the magnitude of such differential biases are
at most 0.2 dex.  We can obtain similar estimates for the sample as a
whole using the fit parameters given in each panel of Figure
\ref{fig:starpops}.  Taking the $3\sigma$ statistical limits on the
slopes of these relations, we find that the differential effects
across the full range of the sample are $\lesssim 0.1$ dex for
$D_n(4000)$, H$\delta$ EW, and restframe color; and $\lesssim 0.15$
dex for H$\alpha$ EW, age, dust extinction, SSFR, and $M_*/L$.

The same is not true using the simple estimate of dynamical mass,
$\msimple$, in place of $M\dyn$.  As might be expected from comparing
Figures \ref{fig:raw} and \ref{fig:corrd}, we find that the
$M_*/\msimple$ relations for different mass bins are largely parallel,
but significantly offset from one another.  Then, because of
correlations between mass and activity, we also find significant
gradients in the $M_*/\msimple$ relations for the sample as a whole;
typically at the 10--$20 \sigma$ level.  The size of differential
biases as a function of all of $D_n(4000)$, H$\delta$ EW, H$\alpha$
EW, age, and SSFR are on the order 0.2--0.4 dex.  We note in
particular that the relatively strong gradient in $\msimple/M_*$ with
H$\alpha$ EW noticed by \citet{DroryBenderHopp} disappears when we use
the structure corrected dynamical mass estimator, $M\dyn$, in place of
the simple estimate $\msimple$; this apparent bias seems to be more
closely linked to structure than to H$\alpha$ emission {\em per se}.
{\em We thus find a very good correspondence between $M_*$ and $M\dyn$
for galaxies in the \citetalias{Guo2009} sample, but only provided we
account for structural and dynamical non-homology.} \looseness-1

\subsection{Color--$M/L$ Relations}

\citet{BellDeJong} have shown that $M_*/L$ and color are strongly
correlated in both the optical and the NIR. This implies that a single
color is enough to make a reasonable estimate of $M_*/L$ \citep[see
  also, \eg, ][]{LonghettiSaracco, GallazziBell}. These kinds of color
relations have since been widely used for high redshift studies. In
\citet{Taylor2009c}, we have shown that the values of $M_*/L$ that we
use here correlate very strongly with $^{0.1}(g-i)$ color; \viz:
\begin{equation}
\log M_*/L_i = -0.82 + 0.83 \times ^{0.1}(g-i) ~ .
\end{equation}
(In this expression, $L_i$ should be understood as referring to the
absolute luminosity in the $^{0.1}i$-band filter; that is, the
$i$-band filter redshifted to $z = 0.1$.  The absolute magnitude of
the sun in the $^{0.1}i$-band is 4.58).  The scatter around this
relation is just 0.03 dex. If we use this relation to predict $M_*/L$
for galaxies in the \citetalias{Guo2009} sample, we again find very
good correspondence between $M_*$ and $M\dyn$ --- in fact, the rms
scatter in $M_*/M\dyn$ is unchanged. \looseness-1

That is, at least from the point of view of consistency between $M_*$
and $M\dyn$, it would seem that $M_*/L$s estimated on the basis of a
single color are not significantly worse than estimates based on full
SED fits.  This is significant because the SEDs that were used to
derive the $M_*/L$s were corrected for emission lines using the SDSS
spectra. This kind of correction is not practicable for, for example,
high redshift studies. Further, we have repeated our analysis using
the $M_*/L$s derived from the SDSS spectra by \citet{Kauffmann2003},
and find similarly good agreement between $M_*$ and $M\dyn$: at least
on average, it would appear that color-derived $M_*/L$s are just as
good as those derived from optical spectroscopy \citep[see
  also][]{GallazziBell}.

\vspace{0.2cm}

As can be seen in Figure \ref{fig:starpops}, we find no statistically
significant trend in $M_*/M\dyn$ with $^{0.1}(g-i)$ color.  Given the
existence of a relation between $M_*/L$ and color, this fact implies a
relation between $M\dyn/L$ and color.  Using the color--$M_*/L_i$
relation given above, the results shown in Figure \ref{fig:starpops}
imply that:
\begin{eqnarray}
\log M\dyn/L_i = && (-0.54 \pm 0.07) \nonumber \\
&& ~~~ + (0.80 \pm 0.05) \times ^{0.1}(g-i) ~ .
\end{eqnarray}
(Again, $L_i$ in this expression should be understood as referring to
the $^{0.1}i$-band.)  For the galaxies in our sample, the scatter
around this relation is 0.14 dex, almost exactly the same as the
scatter around the $M_*$--$M\dyn$ relation.  That is, using only the
$^{0.1}(g-i)$ color, it is possible to predict the dynamical masses of
the galaxies in our sample with a relative uncertainty of $\approx 40
\%$.

\section{Discussion} \label{ch:discussion}


\subsection{(Non)-Homology and Dynamical Mass Estimation}

In retrospect, it is perhaps unsurprising that $\msimple$ is not a
good estimate of dynamical mass. By using a \sersic-fit size and total
magnitude measurements we have allowed for structural non-homology in
our definition of $\msimple$, while at the same time, by using $k = 4$
for all galaxies, we have assumed dynamical homology. In this sense,
the definition of $\msimple$ is not internally consistent.

On the other hand, it is remarkable that the SDSS \model\ sizes can be
used to obtain a reasonably good dynamical mass estimate, under the
assumption of {\em both} structural {\em and} dynamical homology. As
we argue in Appendix \ref{ch:others}, the $n$-dependence of $K_V(n)$
and cthe ovariance between $n$ and the combination $R\eff/L$ are very
nearly equal and opposite to one another. This leaves $M\dyn/L \propto
K_V(n)\sigma^2R\eff/L$ remarkably insensitive to errors in the size
and total magnitude measurements, provided that 1.)  $m_\mathrm{tot}$,
$\Theta\eff$, and $n$ are self-consistently derived, and 2.)\ $M\dyn$
is derived using the appropriate value of $n$ to calculate the
dynamics term $k$ or $K_V(n)$. This means that, for example, assuming
a De Vaucouleurs profile (\ie , $n=4$) to derive $m_r$ and
$\Theta\eff$ for all galaxies ensures that $\msimple/L \approx
M\dyn/L$, provided that the appropriate value of $K_V$ is used; \viz,
$k \approx K_V(n=4) = 4.62$. This point is amply demonstrated in
Figure \ref{fig:others}. That is, $M\dyn/L$ appears to be insensitive
to errors or assumptions in $n$, so long as everything is consistently
derived self-consistently. \looseness-1

\subsection{$K_V(n)$ and Dynamical Mass Estimation}

The inclusion of the term $K_V(n)$ makes the structure-corrected
dynamical mass estimator $M\dyn$ explicitly model dependent.\footnote{By
the same token, the simple dynamical mass estimator $\msimple$ is also
model dependent, inasmuch as it assumes homology, which is patently
wrong.} Further, this prescription for $K_V(n)$ has been derived under
very simple and idealized assumptions (\viz\ a single component,
spherical, and dynamically isotropic distribution), and so can only be
regarded as approximate.

That said, more sophisticated dynamical models can give an indication
as to how large these effects might be.  For the case of anisotropy,
the effects on the value of $K_V(n)$ are on the order of $\lesssim
0.1$ dex, and become less important for larger $n$ \citep[see,
\eg,][]{CiottiLanzoni}.  Further, \citet{Bertin2002} argue that the
galaxy dynamics (or, more accurately, the value of $K_V$) close to the
galaxy center are in principle rather sensitive to the precise shape
of the total mass distribution.  Their results suggest that this
effect is on the order of $\lesssim 0.1$ dex (see their Figure D.1).

In order to probe the dependence of our conclusions on the assumed form
of $K_V(n)$, we have also trialed using an alternate prescription for
$K_V(n)$, given by \citet[][see their Equation 20]{Cappellari2006}. The
main difference between this prescription and the one given in Equation
\ref{eq:kvn} is that it is phrased in terms of the observed velocity
dispersion within the effective radius, $\sigma\eff$, rather than the
central velocity dispersion $\sigma_0$. This prescription thus has a
different dependence on dynamical isotropy and the dark matter profile,
and so provides an indirect means of probing the importance of these
effects. Using the \citet{Cappellari2006} prescription, we find
qualitatively and quantitatively similar results: we find that $M_*
\propto M\dyn^{0.85}$; we still see that $M_*/M\dyn$ depends on $M_*$ at
fixed $n$, and on $n$ at fixed $M_*$ (although this dependence is
somewhat shallower); we see no trends in $M_*/M\dyn$ with apparent
magnitude or redshift, but a weak trend with observed size; and, at fixed
mass, we see no statistically significant trends in $M_*/L$ with stellar
population parameters. \looseness-1

That is, while it is virtually certain that the model used to derive the
prescription for $K_V(n)$ given in Equation \ref{eq:kvn} is wrong in
several important respects, it seems unlikely that accounting for those
effects that are ignored in the model would have a drastic effect on our
results and conclusions. As we have repeatedly stressed, detailed
dynamical modeling is necessary better constrain the `true' values of
$K_V$ for individual galaxies.

\subsection{Comparison to Previous Studies} \label{ch:comp}

\subsubsection{Comparison to Other SDSS Studies}

In comparison to other studies of the relation between stellar and
dynamical mass estimates based on SDSS data, we find considerably less
variation in $M_*/M\dyn$ with mass.  In the case of
\citet{DroryBenderHopp}, this difference is simply due to the fact
that we account for non-homology in the derivation of $M\dyn$; using
the simple mass estimate $\msimple$, we have verified that we are able
to reproduce their results.  In the case of \citet{Gallazzi2006},
there is the additional complication that they use qualitatively
different measures of total flux and size: specifically, the Petrosian
magnitudes and half-light radii given in the basic SDSS catalog, which
are derived directly from the observed curves of growth.  Again, we
have verified that we can reproduce their results using the same
measurements. \looseness-1

\citet{Gallazzi2006} also split their galaxy sample into bins of
\sersic\ index, and find similar slopes to the $M_*$--$\msimple$
relation for each subsample; they find that the logarithmic slope of
the relation varies from 0.847 to 0.801 between $n = 3$ and $n = 5.5$.
The size of this variation is entirely consistent with the results we
have shown in Figure \ref{fig:rawcond}.  They use this fact to argue
that non-homology does not have a significant impact on the slope of
the global $M_*$--$M\dyn$ relation.  However, as we have also shown in
Figure \ref{fig:rawcond}, while the $M_*$--$\msimple$ relations for
each bin in $n$ are parallel, they are significantly offset from one
another.  It is this through this offset, combined with a correlation
between $M_*$ and $n$, that non-homology affects the slope of the
$M_*$--$M\dyn$ relation; \citet{Gallazzi2006} make no mention of such
an offset.  While our conclusions differ with those of
\citet{Gallazzi2006}, our results are thus not obviously inconsistent. \looseness-1

\subsubsection{Comparison to Detailed Dynamical Modeling Results}

\citet{Cappellari2006} have also argued against the idea that
non-homology has an important impact on dynamical mass estimates. This
argument was based on dynamical mass-to-light ratios derived from
detailed 2D and 3D modeling 25 structurally early type galaxies from the
SAURON sample. \citet{Cappellari2006} compared the dynamical
mass-to-light ratios, so derived, to the simple virial mass estimator
$\msimple/L \propto \sigma^2 R_\mathrm{DeV} / L_\mathrm{DeV}$, where
$R_\mathrm{DeV}$ and $L_\mathrm{DeV}$ were derived from De Vaucouleurs
profile fits. They found no evidence for an $n$-dependent offset between
these two quantities. 

The fact that \citet{Cappellari2006} use De
Vaucouleurs-fit sizes and magnitudes is significant: as we have argued
above and in Appendix \ref{ch:others}, when using De Vaucouleurs fits to
derive size and magnitude measurements, it is appropriate to use a
constant $k$ in the definition of $\msimple$. That is, by using De
Vaucouleurs-fit sizes and masses to define $(\msimple/L)$,
\citet{Cappellari2006} effectively guaranteed, almost by
construction, that they would find no structure-dependence when comparing
the `simple' dynamical mass-to-light ratio to that derived from full
dynamical modeling.

\subsubsection{Comparison to Fundamental Plane Studies}

As we have mentioned in the Introduction, the fundamental plane can be
thought of as measuring the variation in the dynamical mass-to-light
ratios of early type galaxies as a function of velocity dispersion,
luminosity, or mass \citep[see, \eg,][]{FP2, Jorgensen1996}. How do our
derived values of $M\dyn$ compare to those derived from the fundamental
plane? To address this question, we selected the non-emission galaxies
from within the \citetalias{Guo2009} sample with $n > 2.5$. For these
galaxies, we find $M\dyn/L \propto \sigma^{0.88 \pm 0.06}$. For
comparison, \citet{Jorgensen1996} find $\sigma^2 R_\mathrm{DeV}
/L_\mathrm{DeV} \propto \sigma^{0.86}$, where again, $R_\mathrm{DeV}$ and
$L_\mathrm{DeV}$ have been derived via De Vaucouleurs fits. Again, the
covariance between the fit values of $m_r$, $\Theta\eff$, and $n$ mean
that $\sigma^2 R_\mathrm{DeV} /L_\mathrm{DeV} \approx M\dyn/L$. Our
structure-corrected dynamical mass estimates are thus in good agreement
with those derived from the fundamental plane. (Parenthetically, we also
note that \citet{Cappellari2006} found that their dynamical mass-to-light
ratios derived from detailed dynamical fits scaled as $\sigma^{0.82}$,
which is also consistent with our results.)

We have also considered how the dynamical-to-stellar mass ratio,
$M\dyn/M_*$ varies with $\sigma$ for this same sample of early type
galaxies: we find $M\dyn/M_* \propto \sigma^{0.50\pm 0.06}$.  This
would suggest that less than half of the tilt of the fundamental plane
is due to variations in the mass-to-light ratios of early type
galaxies as a function of $\sigma$ \citep[\cf , \eg,
][]{PrugnielSimien, Trujillo2004, Allanson2009}.  We present this
result only for completeness; proper interpretation of this result
requires much more detailed analysis, and is beyond the scope of this
work. \looseness-1

We also note that both \citet{PrugnielSimien} and \citet{Trujillo2004}
have made a very similar argument for the importance of non-homology in
estimating dynamical masses as we have made in Sections \ref{ch:results}
and \ref{ch:results2}, based on dynamical mass-to-light ratios derived
from the fundamental plane. Our analysis based on the correspondence
between $M_*$ and $M\dyn$ is complementary to theirs in two ways. First,
their analyses were specific to early type galaxies; we have thus
extended their result to the general galaxy population. Secondly, both
authors focussed on $M_\mathrm{d}/L$, rather than $M_*/M\dyn$; that is,
neither of these authors considered the relation between galaxies'
stellar and dynamical masses.

\subsection{Interpretation}

Turning now to the interpretation of our results, the remarkable
consistency between stellar and dynamical mass estimates shows two
things.  First, it strongly suggests that the measurements of $M_*$
and $M\dyn$ are both meaningful and relatively robust.  In particular,
our results indicate that it is possible to derive stellar mass
estimates without strong differential biases as a function of age,
dust, SSFR, or $M_*/L$, based only on broadband optical photometry (or
indeed on a single optical color).

Secondly, it implies that intrinsic variations in the
stellar-to-dynamical mass ratio (due to, \eg, variations in the
dark--to--stellar mass ratio, variations in the IMF, or dynamical
differences beyond the simple non-homology considered here) as a
function of stellar mass, galaxy structure, and star formation
rate/history are either small, or conspire to leave the inferred
values of $M_*/M\dyn$ relatively unchanged. The rest of this section
will be devoted to discussion of this result.


Since $M\dyn$ is an estimate of total mass, it can only be interpreted
as an upper bound on the true stellar mass. Because we have no {\em a
  priori} means of separating out the relative contributions of
luminous and non-luminous mass to $M\dyn$, the relation between $M_*$
and $M\dyn$ is complicated by degeneracies between the relative
contributions of gas and dark matter, as well as uncertainties in the
low mass shape of the IMF. 

The simplest way to interpret the non-linearity of the $M_*$--$M\dyn$
relation is as indicating a greater central dark matter fraction for
higher mass galaxies, in qualitative agreement with theoretical
expectations. Using simple arguments based on the observed dynamics of
elliptical galaxies, \citet{Franx1993} and \citet{Kochanek1994} have
argued that accounting for a dark matter halo implies $|\log M_*/M\dyn|
\sim 0.14$--0.18 dex. This would go a long way towards explaining the
$\sim 0.23 \pm 0.03$ offset that we have observed. \looseness-1

We can estimate gas masses using the prescription given by
\citet{Zhang2009}. These authors have used a sample of relatively low
mass SDSS galaxies with literature H{\footnotesize I} masses to derive a
prescription for $M_\mathrm{HI}/M_*$ as a function of $(g-r)$ color and
stellar surface density. Using this prescription to derive baryonic mass
estimates, $M_\mathrm{bar} = M_* + M_\mathrm{HI}$, reduces the size of
the offset between $M_\mathrm{bar}$ and $M\dyn$ by 0.05 dex to $-0.18$
dex, and brings the logarithmic slope of the $M_\mathrm{bar}/M\dyn$
relation to 0.95. The fact that the \citet{Zhang2009} relation has been
derived for very different galaxies to the ones we consider here means
that this result should be interpreted with caution. Even so, it is
striking that, taken together, the estimated contributions of
H{\footnotesize I} and dark matter almost perfectly explain the observed
offset between $M_*$ and $M\dyn$, and imply only a mild trend in
$M_\mathrm{bar}/M\dyn$ with mass: $M_\mathrm{bar}/M\dyn \propto
M_\mathrm{bar}^{-0.05}$.

Then there is the matter of the IMF. The effect of adopting a
\citet{Salpeter} IMF rather than that of \citet{Chabrier2003} would be
approximately to scale all our values of $M_*$ up by 0.22 dex. For a
linear $M_*$--$M\dyn$ relation (which our data are only marginally
consistent with), this would leave virtually no room for dark matter or
gas in the centers of galaxies in our sample. For the slightly
less-than-linear relation preferred by our data, this would imply that
$M_* > M\dyn$ for galaxies with $M_* \lesssim 10^{11}$ M\sol, which is
logically inconsistent. Thus we can say that, at best, our results are
only marginally consistent with a \citet{Salpeter} IMF. Accounting for
dark matter, our results are also weakly inconsistent with a `diet
Salpeter' IMF, and completely consistent with a \citet{Kroupa2001} or
\citet{Chabrier2003} IMF. (Parenthetically, we also note that the results
shown in Figure \ref{fig:starpops} can also provide a weak constraint on
variations in the IMF as a function of star formation rate/history.)
\looseness-1

Finally, we note that the observed scatter around the $M_*$--$M\dyn$
relation is rather small: just 0.13 dex. We argue in Appendix
\ref{ch:others} that the ratio $M_*/M\dyn \propto \sigma^2 R\eff / L$
is remarkably insensitive to errors in the \sersic -fit parameters,
provided that they are consistently derived, and that dynamical
non-homology is taken into account. This implies that the
uncertainties in $M_*/M\dyn$ are dominated by errors in the
measurement of $\sigma_0$ and $M_*/L$. The mean formal uncertainty in
$\sigma$ for our galaxy sample is 0.034 dex.  We estimate the mean
random error in $M_*/L$ to be on the order of 0.1 dex; this is the
random scatter between the SED-fit $M_*/L$s used here and the
spectrally derived $M_*/L$s given by
\citet{Kauffmann2003}.\footnote{Note that this is almost certainly an
  underestimate of the `true' random uncertainty in $M_*/L$. Including
  NIR data (where the stellar population models are the most
  uncertain), and properly accounting for propagation of uncertainties
  in stellar population models and the IMF, \citet{Conroy} argue that
  the uncertainties in $M_*/L$ are on the order of 0.3 dex.} Adding
these errors in quadrature (\ie\ neglecting correlations between
$\sigma$ and $M_*/L$ at fixed $M_*$) produces an uncertainty in
$M_*/M\dyn$ of 0.12 dex. This would imply that, at fixed $M_*$, the
intrinsic scatter in $M_*/M\dyn$ is potentially very small indeed:
$\lesssim 0.04$ dex.  \looseness-1

\section{Summary} \label{ch:summary}

The central focus of this work has been the degree of consistency
between stellar and dynamical mass estimates, based on the latest
generation of data products from the SDSS.  We have shown that
structural differences in galaxy dynamics can have a large impact on
the estimated values of dynamical mass, and so on the degree of
correspondence between stellar and dynamical mass (Section
\ref{ch:results}; Figures \ref{fig:raw} and \ref{fig:corrd}).
Provided we account for structure-dependent differences in galaxy
dynamics (using the term $K_V(n)$, as defined in Equations
\ref{eq:mdyn} and \ref{eq:kvn}), we find very good agreement between
the inferred stellar and dynamical masses of galaxies within the
\citetalias{Guo2009} sample. \looseness-1

Our analysis is based on the carefully-constructed satellite/central
galaxy sample of \citetalias{Guo2009}, making selection effects a major
potential concern. However, we find no signs of major differences in the
relation between $M_*$ and $M\dyn$ for central/satellite or non-/emission
galaxies within the sample, suggesting that our results are not seriously
affected by selection effects (Section \ref{ch:selfx}; Figure
\ref{fig:subsamples}). Moreover, we find qualitatively and quantitatively
similar results analyzing a more general sample of $0.035 < z < 0.08$
galaxies, using the best-fit \sersic\ parameters given in the NYU VAGC,
or using the De Vaucouleurs/exponential \model\ fit parameters given in
the basic SDSS catalog (Appendix \ref{ch:others}). \looseness-1

We find that the ratio $M_*/M\dyn$ varies with both $M_*$ and $n$
(Sections \ref{ch:structure} and \ref{ch:mstarn}; Figures
\ref{fig:rawcond} and \ref{fig:corrcond}). While the apparent
$n$-dependence of $M_*/M\dyn$ is sensitive to the assumed form of
$K_V(n)$, changing $K_V(n)$ cannot affect the result that $M_*/M\dyn$
varies with $M_*$ at fixed $n$. Without spatially resolved dynamical
information for individual galaxies, however, we cannot determine whether
the apparent mass-dependence of $M_*/M\dyn$ is caused by some
mass-dependent difference in galaxy dynamics, rather than a genuine
physical difference in the stellar-to-dynamical mass ratios of galaxies
with different masses. \looseness-1

Similarly, while we have shown very good agreement between stellar and
dynamical masses for SDSS galaxies, we cannot unambiguously prove that
neither of these quantities suffers from systematic biases. On the other
hand, using the \sersic -fit parameters given by \citetalias{Guo2009}, we
do not see any systematic variation in $M_*/M\dyn$ with observed
properties like apparent magnitude, apparent size, or redshift (Section
\ref{ch:bias}; Figure \ref{fig:bias}). This is not true if we use the
\sersic -fit parameters given in the NYU VAGC, which has been shown to
suffer from systematic errors arising due to background over-subtraction.
That is, we have the ability to detect these sorts of errors, and do not
see evidence for such errors for our sample. Further, we do not see any
signs of variation in $M_*/M\dyn$ for galaxies with different stellar
populations, or for galaxies in different states of activity (\ie , AGN
hosts, star forming galaxies, or non-emission galaxies).

These results, together with the good general agreement between $M_*$ and
$M\dyn$ provide strong circumstantial evidence (but not proof beyond a
reasonable doubt) that there are no serious systematic biases in the
values of $M_*$ and $M\dyn$ that we use here. This implies that the
assumption of non-homology gives the wrong dynamical mass. Further, this
suggests that there are not strong biases in the $M_*/L$s we have used
here: at 99 \% confidence, the consistency between $M_*$ and $M\dyn$
implies any differential biases in the estimate of $M_*/L$ across a wide
range of stellar populations are at the level of $\lesssim 0.12$ dex
($\approx 40$ \%). \looseness-1

\vspace{0.2cm}

\textbf{Acknowledgments.} This work was supported through grants by the
Nederlandse Organisatie voor Wetenschappelijk Onderzoek (NWO), the Leids
Kerkhoven-Bosscha Fonds (LKBF). \\

\clearpage

\begin{appendix}

\section{Validating the SDSS DR7 \\ Velocity Dispersion Measurements} \label{ch:faber}

\begin{figure*} \centering
\includegraphics[width=16.7cm]{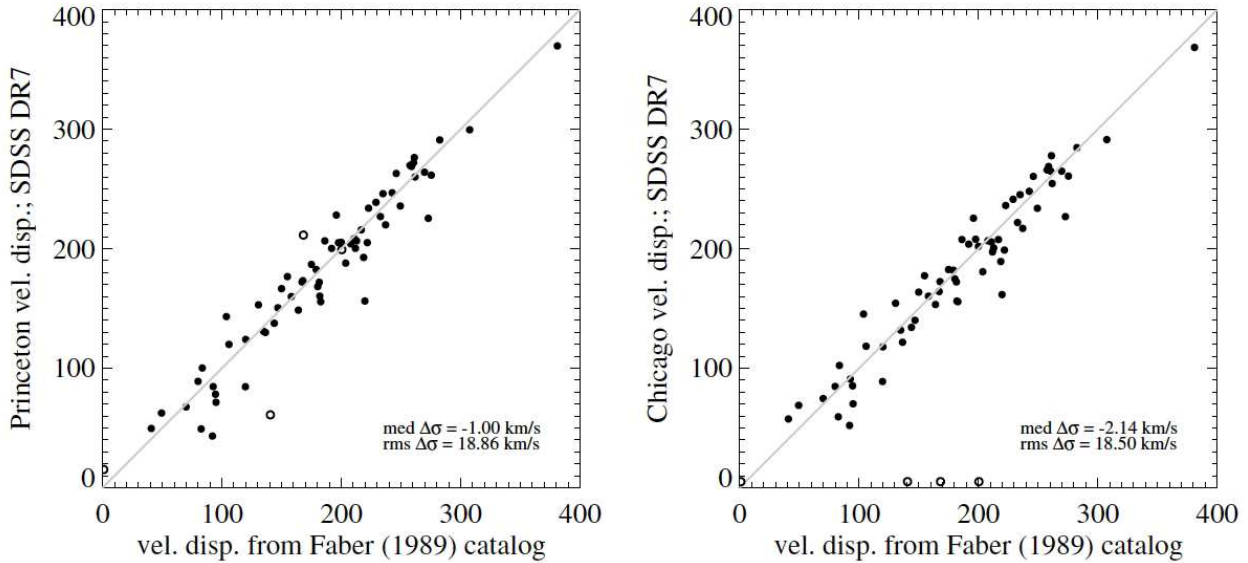}
\caption{Comparison between the two different velocity dispersions
  measurements given in the SDSS DR7 catalog and those given by
  \citet{Faber}. --- \citet{Bernardi2007} have shown that the DR5 SDSS
  velocity dispersions suffered from systematic biases in comparison
  to those from the ENEAR sample as well as earlier SDSS releases.
  For this reason, the algorithms for estimating velocity dispersions
  from SDSS spectra were substantially revised for DR6 and later.  In
  each panel of this Figure, we compare one of the two SDSS velocity
  dispersion measurements to those in the `seven samurai' catalog
  \citep{Faber}.  There are no signs of any systematic problems with
  the DR7 SDSS velocity dispersions.  (Note however that the scatter
  in these comparisons is significantly higher than would be expected
  from the formal measurement uncertainties, which are on the order of
  $3.5$ km/s.) \label{fig:faber}}
\end{figure*}

\citet{Bernardi2007} showed that there was an inconsistency between the
$\sigma$--$L$ relations for early type galaxies derived using the early
data release (EDR) and DR5 SDSS catalogs. Further, she was able to show
that the cause for this discrepancy was systematic biases in the DR5
velocity dispersions: in comparison to literature values from HyperLeda,
the DR5 measurements slightly but systematically over-estimated the
velocity dispersions of intrinsically low-$\sigma$ galaxies. For DR6+,
partially in response to the findings of \citet{Bernardi2007}, the SDSS
velocity dispersion pipelines were substantially revised. The new
dispersions have been shown to agree well with the EDR velocity
dispersions used by \citet{Bernardi2003a, Bernardi2003b}, and thus, by
implication, with the improved estimates for DR5 derived by
\citet{Bernardi2007}.\footnote{See
http://www.sdss.org/dr7/algorithms/veldisp.html}

In this Appendix, in order to validate the DR7 velocity dispersions,
we present a comparison between the velocity dispersions given in the
basic SDSS DR7 catalog to those given by \citet{Faber} for elliptical
galaxies in their sample.  The results of this comparison are shown in
\textbf{Figure \ref{fig:faber}}.  The left panel of this Figure shows the
comparison for the Princeton or SpecBS values of $\sigma$; the right
panel shows that for the Chicago or spectro1d values of $\sigma$.
Note that the Chicago algorithm only outputs values of $\sigma$ for
those galaxies that are spectroscopically classified as being early
type; the three \citet{Faber} galaxies at the bottom of the right-hand
panel are not classified as being early type, and so are not given
Chicago velocity dispersions.

Within both panels, we give the median and rms difference between the
SDSS and \citet{Faber} velocity dispersion measurements.  It is clear
from this Figure that neither of the DR7 velocity dispersions suffers
from serious systematic biases in comparison to the \citet{Faber}
measurements.  We note, however, that the rms scatter, which is on the
order of 19 km/s, is considerably higher than the median formal
measurement uncertainty given in the SDSS catalog, which is on the
order of 3.5 km/s for the galaxies shown in Figure \ref{fig:faber}.
That is, it seems that the formal uncertainties on the SDSS velocity
dispersions significantly underestimates the true error, at least for
the relatively bright galaxies shown here. \looseness-1

\clearpage

\section[Repeating our Analysis for a General Galaxy Sample]
{Selection Effects and Systematic Biases: Repeating our Analysis for a General Galaxy Sample} \label{ch:others}

As we have repeatedly stressed in the main text, the \citetalias{Guo2009}
sample that we analyze in the main text is heavily selected.  In order
to make sure that our conclusions are not unique to the
\citetalias{Guo2009} sample, in this Appendix we repeat our analysis for a
more general galaxy sample.  For this exercise, we have selected
$m_{\mathrm{Pet},r} < 17.5$ galaxies with \texttt{sciencePrimary} spectra
in the range $0.035 < z < 0.08$.  As in our main analysis, we also
require that the relative error on the velocity dispersion is less
than 10 \%, and that $\sigma\obs > 75$ km/s; these selections
effectively limit the sample to $M_* \gtrsim 10^{10}$ M\sol .  The
additional incompleteness due to our velocity dispersion criteria is
less than 10 \% for all $M_* > 10^{10.3}$ M\sol\ and $n \gtrsim 3$,
but is significant for $n \lesssim 1$ at all masses.

There is one complicating factor in the comparison between this general
field sample and the \citetalias{Guo2009} sample that we discuss in the
main text. For the field sample, we are forced to rely on either the De
Vaucouleurs/exponential \model\ fits provided in the basic SDSS catalog,
or the \sersic\ fits given in the NYU VAGC. (Recall that we need a measure
of total magnitude to derive $M_*$, and both an effective radius and a
\sersic\ index measurement to derive $M\dyn$.) Both of these sets of
measurements have their faults. The SDSS \model\ fits are overly
simplistic in that they assume that $n$ is equal to either 1 or 4; this
will clearly introduce systematic errors in the fit quantities as a
function of (intrinsic) profile shape. The VAGC \sersic\ fits are also
known to suffer systematic errors \citep{vagc}, due to background
over-subtraction \citepalias{Guo2009}.

\vspace{0.2cm}

\begin{figure*} \centering
\includegraphics[width=16.7cm]{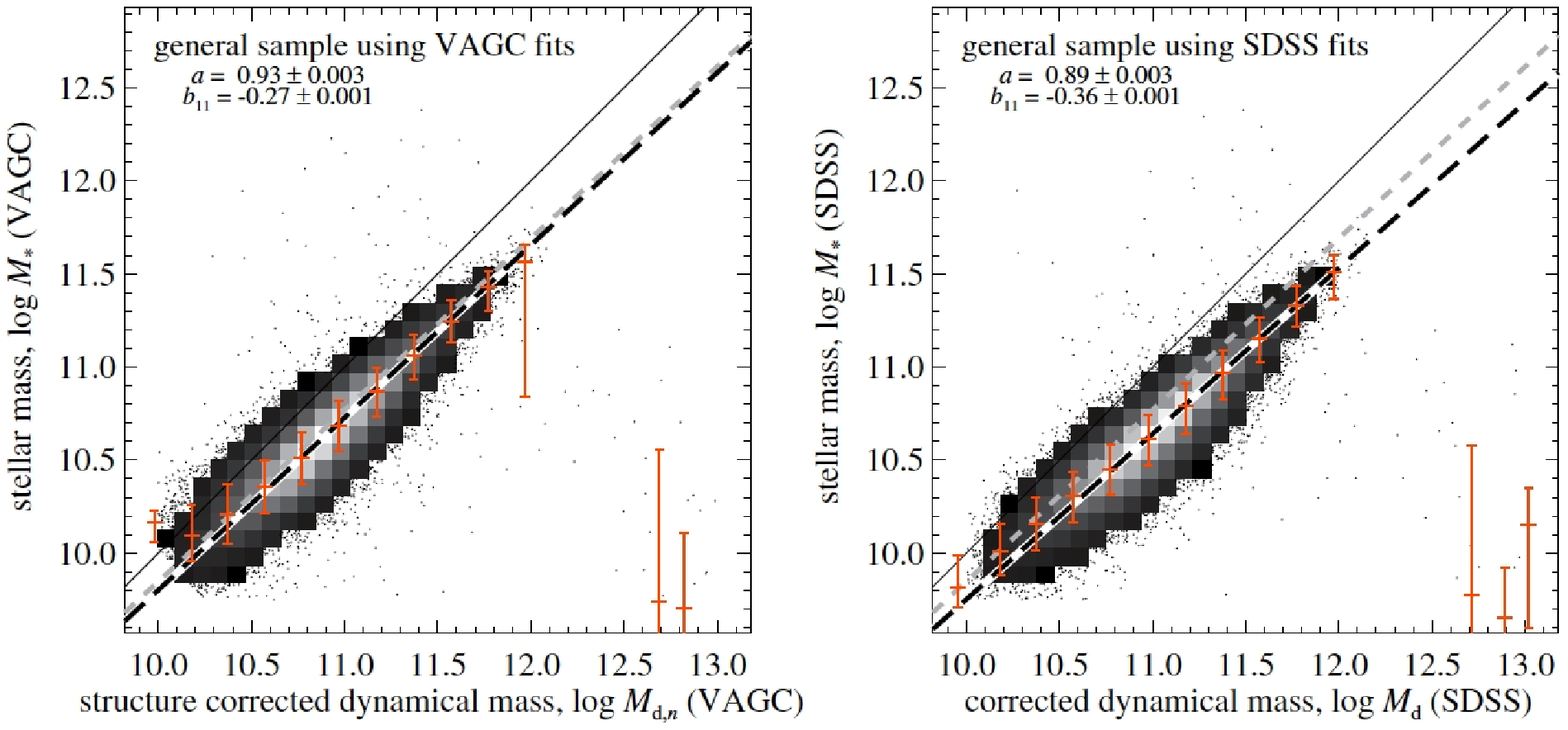}
\caption{Comparing dynamical and stellar mass for a general galaxy
  sample. --- For this Figure, we have selected $0.035 < z < 0.08$
  galaxies with \texttt{sciencePrimary} spectra, $\sigma\obs > 75$
  km/s, and $\Delta\sigma\obs/\sigma\obs < 0.1$.  For each panel, we
  have derived $M\dyn$ using either \sersic\ structural parameters
  from the NYU VAGC \citep[][left panel]{vagc}, or using De
  Vaucouleurs/exponential \model\ structural parameters from the basic SDSS
  catalog (right panel).  As in other Figures, the solid lines show
  fits to the data; the points with error bars show the median
  relations in bins.  The grey dashed line shows the $M_*$--$M\dyn$
  relation we derive for the \citetalias{Guo2009} catalog, using their
  \sersic\ structural parameters.  The general $M_*$--$M\dyn$ relation
  for field galaxies is very similar to the one we find for the
  heavily-selected \citetalias{Guo2009} sample. \label{fig:others}}
\end{figure*}
\begin{figure*} \centering
\includegraphics[width=16.7cm]{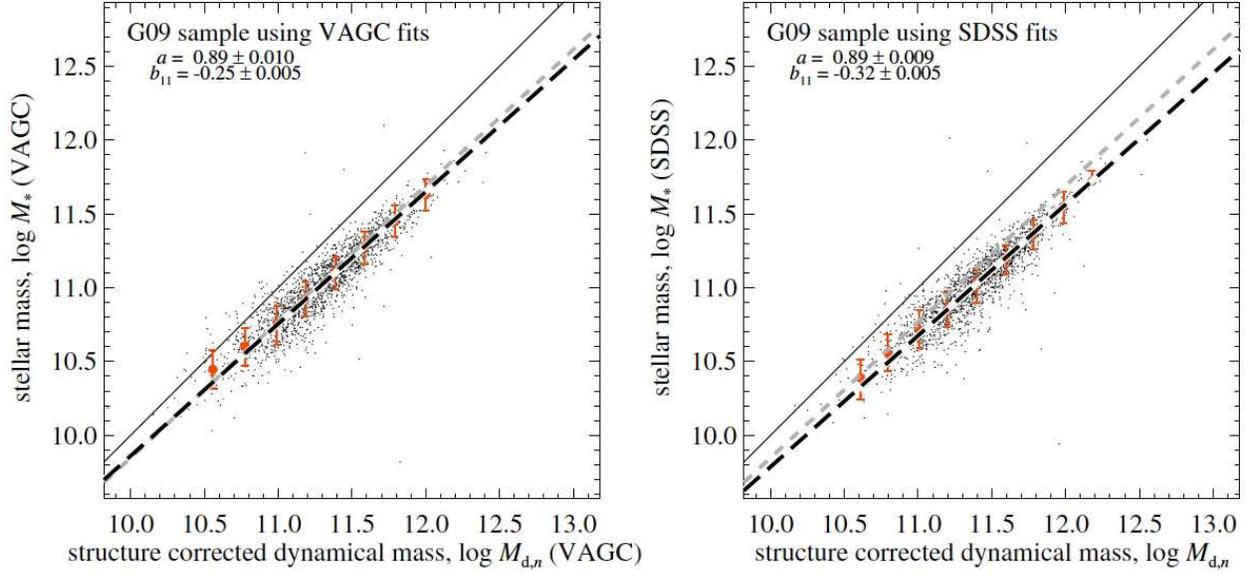}
\caption{Comparing dynamical and stellar mass for the \citetalias{Guo2009}
  sample, using structural fit parameters from the NYU VAGC or the
  basic SDSS catalog. --- All symbols and their meanings are directly
  analogous to Figures \ref{fig:corrd} and \ref{fig:others}.  In
  comparison to Figure \ref{fig:corrd}, this Figure differs only in
  that we have used the \sersic-fit parameters from either the NYU
  VAGC \citep{vagc} or the De Vaucouleurs/exponential \model\ fit parameters
  from the basic SDSS catalog.  The fact that these results agree very
  well with those shown in Figure \ref{fig:corrd} shows that the ratio
  $M_*/M\dyn$ is not extremely sensitive to errors in the structural
  fit parameters.  In comparison to Figure \ref{fig:others}, the
  difference is that we only show galaxies in the \citetalias{Guo2009}
  sample.  The fact that the results in this Figure agree very well
  with those shown in Figure \ref{fig:others} show that sample
  selection effects do not play an important role in our
  results. \label{fig:measures}}
\end{figure*}

Bearing both these issues in mind, in \textbf{Figure \ref{fig:others} }we show the
relation between $M_*$ and $M\dyn$ for our general, field galaxy sample,
using either the \sersic\ fit parameters from the NYU VAGC (left panel) or
the \model\ fit parameters from the basic SDSS catalog (right panel).
Using either set of parameters, the results for this general sample agree
really very well with what we have found for the \citetalias{Guo2009}
sample in Figure \ref{fig:corrd}. Moreover, the two panels in Figure
\ref{fig:others} agree remarkably well with one another, even despite the
significant and very different systematic errors that each set of
measurements suffers from.

How can this be?  It turns out that the covariance between \sersic\
parameters leaves ratio $M_*/M\dyn$ is remarkably robust to both
random and systematic errors in the \sersic\ fits, provided $M\dyn$ is
calculated as per Equation \ref{eq:mdyn}.  To illustrate this, let us
compare the \citetalias{Guo2009} and VAGC measurements.  Although there are
large differences in all three parameters individually, there are
tight correlations between $\Delta n$, $\Delta m_\mathrm{tot}$, and
$\Delta R\eff$.  (Here and in what follows the `$\Delta$' implies the
difference between the VAGC and \cite{Guo2009}-derived value, in the
sense of VAGC-minus-\citetalias{Guo2009}.)  Now, $M_*$ scales directly with
total flux; fitting to $\Delta M_*$ as a function of $\Delta n$, we
find that $\Delta\log M_* \propto 0.04 \Delta n$, with an rms scatter
in $\Delta \log M_*$ of 0.07 dex.  $M\dyn$, at least as defined in
Equation \ref{eq:mdyn}, depends on both the effective radius and
\sersic\ index.  Considering the change in $M\dyn$ due to changes in
size alone, we find $\Delta \log M\dyn \propto 0.11 \Delta n$, with a
scatter of 0.07 dex; for the effect due to changes in the \sersic\
index alone, we find $\Delta \log M\dyn \propto -0.06 \Delta n$, with
an rms scatter of 0.03 dex.  Taken together, the overall change in
$M\dyn$ scales with $\Delta n$ as $\Delta \log M\dyn \propto 0.04
\Delta n$.

Thus we see that the changes in $M_*$ and $M\dyn$ thus have virtually the
same dependence on $\Delta n$, leaving the ratio $M_*/M\dyn$ virtually
unchanged. Further, the scatter in $\Delta(M_*/M\dyn)$ is just 0.04 dex.
Using the basic SDSS \model\ fits, while we find slightly stronger
dependences with $\Delta n$, we still find that the ratio $M_*/M\dyn$
remains very robust. We stress that the above argument only holds if we
account for the dynamical effects of structure in the calculation of
$M\dyn$: if we removed the $n$-dependence of $M\dyn$ that enters via
$K_V(n)$, then we would find that the ratio $\Delta\log( M_*/\msimple)
\propto -0.06 \Delta n$, in agreement with the expectation from the
analysis immediately above. \looseness-1

\vspace{0.2cm}

To explicitly demonstrate that the observed relation between $M_*$ and
$M\dyn$ is not particularly sensitive to the measurements used to derive
the values of $M_*$ and $M\dyn$, in \textbf{Figure \ref{fig:measures}} we
show the $M_*$--$M\dyn$ relation for the \citetalias{Guo2009} sample
analyzed using structural parameters from the NYU VAGC (left panel) or
from the SDSS catalog (right panel). In comparison to Figure
\ref{fig:corrd}, the slope of the $M_*$--$M\dyn$ relation for the
\citetalias{Guo2009} sample is very similar using any of the three sets of
structural parameters: 0.92 for the \citetalias{Guo2009} fits, compared to
0.89 for the NYU VAGC fits, and 0.88 for the SDSS \model\ fits. The
normalization of the $M_*$--$M\dyn$ relation is slightly more sensitive:
at $M\dyn = 10^{11}$ M\sol, we find that $\Delta \log (M_*/M\dyn)$ =
$-0.23$, $-0.24$, and $-0.32$ dex using the \citetalias{Guo2009}, VAGC,
and SDSS fits, respectively.  

In comparison to Figure \ref{fig:others}, the results in Figure
\ref{fig:measures} also demonstrate that the $M_*$--$M\dyn$ relation
for the \citetalias{Guo2009} is very similar to that for a more general
field galaxy sample.  For example, using structural parameters from
the VAGC, the logarithmic slope and intercept of the $M_*$--$M\dyn$
relation are $a = 0.89$ and $b_{11} = -0.24$ for the \citetalias{Guo2009}
sample, compared to $a = 0.91$ and $b_{11} = -0.27$ for the general
galaxy sample.  

\vspace{0.2cm}

\begin{figure*} \centering
\includegraphics[width=12cm]{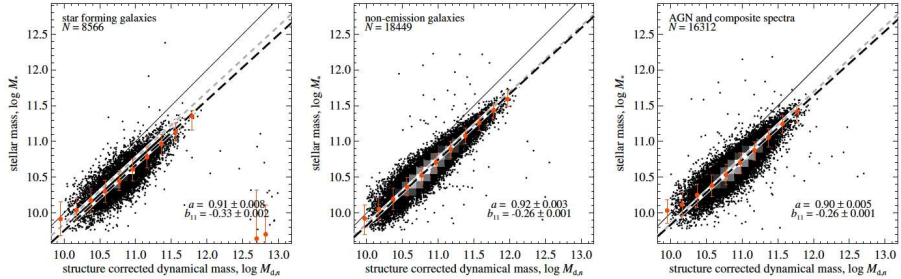}
\caption{The $M_*$--$M\dyn$ relation for galaxies in different states
  of activity. --- Each panel of this Figure shows the relation between
  $M_*$ and $M\dyn$ for different subsamples of the general $0.035 < z
  < 0.08$ galaxy population, split according to their spectral
  classification, and analyzed using the \sersic\ fits given in the
  NYU VAGC.  From left to right, we show non-emission galaxies, star
  forming galaxies, and AGN/composite spectra galaxies; the spectral
  classifications are those of \citet{Brinchmann}, which are based
  on the BPT diagram.  In each panel, the heavy dashed line shows the
  best fit $M_*$--$M\dyn$ relation; for comparison, the grey
  short-dashed line shows the best fit relation for the sample as a
  whole.  While each subsample follows a similar $M_*$--$M\dyn$
  relation, there is an offset between the different relations, on the
  order of 0.07 dex.  As for the \citetalias{Guo2009} sample, the differences
  between the different subsamples disappears if we consider only $n >
  3$ galaxies.  That is, these offsets appear to be due to the
  different distributions of $n$ within each subsample, rather than
  intrinsic differences in the values of the stellar-to-dynamical mass
  ratio for galaxies in different states of activity.  This argues the
  idea that selection effects play a major role in shaping our
  results. \label{fig:othersubs}}
\end{figure*}

\begin{figure*} \centering \vspace{-0.1cm}
\includegraphics[width=16.8cm]{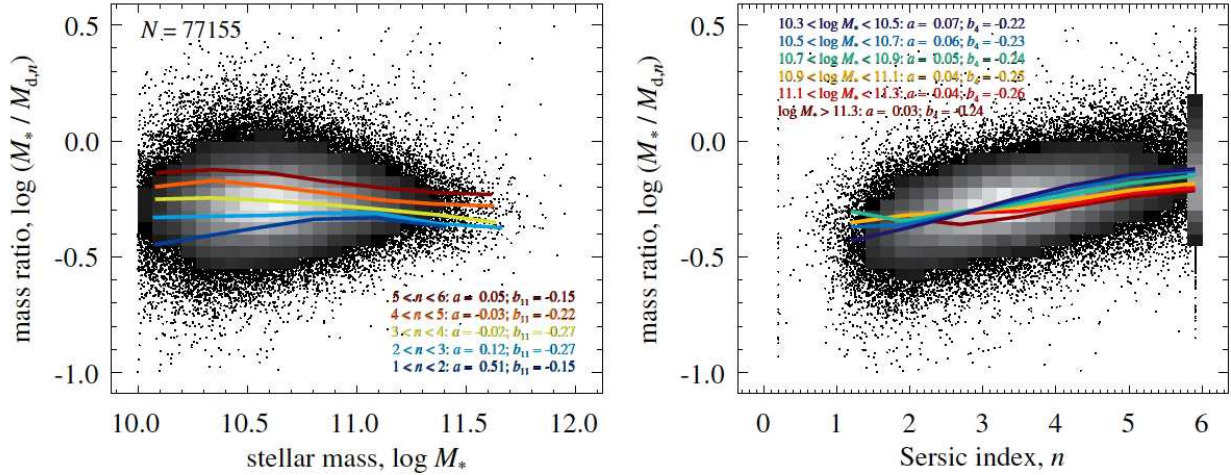}
\caption{Separating out the mass- and structure-dependence of the mass
  ratio $M_*/M\dyn$ of a field sample of $0.035 < z < 0.08$ galaxies,
  showing active and passive galaxies separately. --- In analogy to
  Figure \ref{fig:corrcond}, the colored lines in the left panels show
  the mass-dependence of the mass ratio $M_*/M\dyn$ in bins of
  \sersic\ index; those in the right panel shows how $M_*/M\dyn$
  varies with $n$ in bins of $M_*$.  The precise bins are given within
  each panel, along with the best fit parameters for the relation for
  each bin.  The results in this panel have been derived using the NYU
  VAGC \sersic\ fit parameters.  In comparison to Figure
  \ref{fig:corrcond}, we see qualitatively similar behavior for
  $M_*/M\dyn$ as a function of both $M_*$ and $n$ for the general
  galaxy sample as we do for the heavily-selected \citetalias{Guo2009}
  sample.  Moreover, we point out that this is true for non-emission,
  star forming, AGN and composite spectra galaxies separately, as well
  as for the general sample as a whole.  Because the NYU fits suffer
  systematic biases, there are quantitative differences in the results
  shown in this Figure and those in Figure \ref{fig:corrcond}.
  Despite these quantitative differences, the relatively weak
  dependence of $M_*/M\dyn$ on $n$ for each bin in $M_*$ supports our
  main result; \viz, that accounting for non-homology leads to
  reasonably good consistency between stellar and dynamical mass
  estimates.
\label{fig:othercond}}
\end{figure*}

In \textbf{Figure \ref{fig:othersubs}}, we show that galaxies in different
states of activity follow very similar $M_*$--$M\dyn$ relations.  In
this Figure, we have split the general galaxy sample into
non-emission, star forming, and AGN and composite spectra subsamples
using the \citet{Brinchmann} spectral classification scheme described
in Section \ref{ch:selfx}).  The logarithmic slopes of the
$M_*$--$M\dyn$ relation for each subsample agree with one another, as
well as with that for the sample as a whole, to within a few percent.
We do find that the $M_*$--$M\dyn$ relation for the star forming
subsample is offset from that for the non-emission and AGN/composite
subsamples, at the level of 0.07 dex.  However, as for the
\citetalias{Guo2009} sample (see Section \ref{ch:selfx}), these small
differences disappear if we consider only $n > 2.5$ galaxies.  We thus
conclude that these differences are principally driven by the
different distribution of $n$ values within the star forming sample,
rather than intrinsic differences in the stellar-to-dynamical mass
ratios of star forming galaxies.

Finally, in \textbf{Figure \ref{fig:othercond}}, we separate out the
$M_*$- and $n$-dependences of $M_*/M\dyn$ for the general galaxy sample,
analyzed using the NYU VAGC \sersic -fit parameters. Again, we find that
the ratio $M_*/M\dyn$ depends on both $M_*$ (at fixed $n$) and on $n$ (at
fixed $M_*$). The results in this Figure suggest that the mass-dependence
of $M_*/M\dyn$ may flatten considerably for $n \lesssim 2$ and $10
\lesssim \log M_*/ $M\sol\ $\lesssim 10.5$ (\ie\ below the mass limit of
the \citetalias{Guo2009} sample). 

As we have noted above, there are
significant differences in the values of $M_*$ and $n$ given by
\citetalias{Guo2009} and in the VAGC. These differences in $M_*$ and $n$
mean that the results in this Figure are not in quantitative agreement
with those shown in Figure \ref{fig:othercond}, even though both datasets
show good agreement in the global $M_*$--$M\dyn$ relation. In particular,
the slope of the $M_*/M\dyn$--$n$ relation at fixed $M_*$ is significantly
steeper than we find for the \citetalias{Guo2009} sample. This is at least
partially due to the bias in the NYU values of $n$; the NYU values are
systematically lower than the \citetalias{Guo2009} values, which has the
effect of steepening the $M_*/M\dyn$--$n$ relation.

With these caveats, the main conclusion to be drawn from Figure
\ref{fig:othercond} is that accounting for non-homology in the
derivation of dynamical masses leads to considerably better
consistency between $M_*$ and $M\dyn$ (as a function of $n$, and at
fixed $M_*$), in agreement with our findings in Section
\ref{ch:results2}.  Further, we note that we find similar and
consistent behavior in $M_*/M\dyn$ at fixed $M_*$ and $n$ for each of
the three subsamples shown in Figure \ref{fig:othersubs}, in agreement
with our conclusions above.

\vspace{0.2cm}

In summary, then, in this Appendix we have demonstrated two things.
First, we have shown that we find very similar results for the
\citetalias{Guo2009} sample, analyzed using the results of the \sersic\
fits given by \citetalias{Guo2009}, and for a more general galaxy sample,
analyzed using either the \sersic\ fits given in the NYU VAGC or the
De Vaucouleurs/exponential \model\ fits given in the basic SDSS catalog.
Secondly, we have shown that we find very similar results for the
\citetalias{Guo2009} sample analyzed using any of these three sets of
structural parameters.  The most important conclusion to be drawn from
these results is that the results we have presented in the main text
are not driven, nor particularly sensitive to, selection effects. \looseness-1

\end{appendix}

\end{document}